\documentclass[reprint,superscriptaddress,amsmath,amssymb,aps,prb,floatfix]{revtex4-1}
\usepackage[T1]{fontenc}
\usepackage{upgreek}
\usepackage{lmodern}
\usepackage{xcolor}
\usepackage{multirow}
\usepackage{graphicx}
\usepackage{subfig}
\DeclareGraphicsExtensions{.png,.pdf,.eps}
\graphicspath{{figures/}}

\newcommand{\degree}{\ensuremath{^\circ}}

\begin{document}
	
	\preprint{APS/123-QED}
	
	\title{\emph{Ab initio} theory of graphene-iron(II) phthalocyanine hybrid systems as scalable molecular spintronics}
	
	\author{Marcin Roland Zem{\l}a}
    \email[]{Marcin.Zemla@pw.edu.pl}
    \affiliation{Materials Design Division, Faculty of Materials Science and Engineering, Warsaw University of Technology, Wo{\l}oska 141, 02-507 Warsaw, Poland}
    \affiliation{Institute of Theoretical Physics, Faculty of Physics, University of Warsaw, Pasteura 5, 02-093 Warsaw, Poland}
	\author{Kamil Czelej}
    \affiliation{Institute of Theoretical Physics, Faculty of Physics, University of Warsaw, Pasteura 5, 02-093 Warsaw, Poland}
    \author{Jacek A. Majewski}
    \affiliation{Institute of Theoretical Physics, Faculty of Physics, University of Warsaw, Pasteura 5, 02-093 Warsaw, Poland}

	\begin{abstract}
	
    Graphene - transition metal phthalocyanine (G-MPc) hybrid systems constitute promising platforms for densely-packed single-molecule magnets (SMMs). Here, we selected iron(II) phthalocyanine (FePc) and investigated its interaction with pristine and defective graphene layers employing density functional theory. Our calculations indicate that thorough proper dehydrogenation of the benzol rings in the FePc molecule its adsorption to graphene is thermodynamically favorable. In general, the presence of anchoring sites on the graphene layer, \emph{i.e.} point defects, additionally facilitates the adsorption of FePc, allowing one to achieve high density of SMMs per unit area. Using the combination of group theory, ligand field splitting, and the calculated PBE0 Kohn-Sham eigenvalue spectrum, we resolved the electronic structure and predicted the spin states of both, the isolated FePc and G-FePc hybrid systems. Regardless of adsorption site and the number of removed hydrogen atoms from the benzol rings of FePc, the magnetic moment of the SMM remains unchanged with respect to free FePc. These results should mediate a successful synthesis of densely-packed G-MPc systems and may open up new avenue in designing scalable graphene - SMMs systems for spintronics applications.%
	
	\end{abstract}
	\maketitle

	\section{Introduction}%
    Magnetic molecular nanomaterials have attracted a great deal of attention due to their potential application in spintronic devices, optoelectronics, and the emerging field of quantum information processing\cite{01Cornia2017,02Cinchetti2017,03Yu2014,04Warner2013}. The major problem associated with magnetic nanomaterials, however, arises from random distribution of the easy magnetization axis, which is directly related to the magnetic anisotropy energy (MAE). It has been demonstrated that MAE can be substantially increased by proper engineering of magnetic molecules, specifically by modifying their size, shape, and substrate they are grafted to \cite{05Gambardella2002,06PhysRevLett.117.030802}. %

    Amongst a variety of molecular nanomaterials, the single molecule magnets (SMMs) have intrinsically high value of MAE \cite{07Gatteschi2000,08Mannini2009,09Mannini2005}. SMMs display very interesting properties, such as slow magnetic relaxation \cite{10Sessoli1993,11Woodruff2013} and quantum tunneling of magnetization (QTM) \cite{12Gatteschi2003,13Thomas1996,14Friedman1996}. High value of MAE in SMMs originates from high activation barrier for the spin flip transitions mediated by spin-orbit coupling (SOC). The energy barrier that prevents spontaneous reorientation of the electron-spin in SMMs is proportional to the product $|D|\cdot{S^2}$, where $D$~is the zero-field splitting term and $S$ is the total spin value in the ground state \cite{12Gatteschi2003}. Progress in the molecular design and synthesis techniques has enabled the improvement of working temperature of SMMs \cite{14Friedman1996}. Although high value of MAE in SMMs can significantly hamper the thermal spin flipping, the QTM phenomenon is another mechanism responsible for spin flip transitions \cite{15Ishikawa2005,16Ganzhorn2016,17Urdampilleta2011}. During the QTM, two degenerate electronic states having opposite spin orientations tunnel into each other through the potential energy surface (PES). As a result, the magnetic hysteresis exhibits no gap, which is similar to paramagnetic materials. It has been shown that QTM can be significantly reduced by immobilizing SMMs onto carbon nanotubes \cite{16Ganzhorn2016,17Urdampilleta2011,18Ganzhorn2013}, MgO substrate \cite{19Wackerlin2016}, or the adsorption on Co bilayer islands formed on Au(111) surface \cite{20Ara2019}. %

    Metal phthalocyanines (MPcs), commonly used as dyes and pigments, are supposedly the best candidates for deposition of thin films under ultra-high vacuum (UHV) conditions due to: (i) their ideal thermal and chemical stability when used in the molecular beam techniques, and (ii) the tendency to grow in ordered phases \cite{21Papageorgiou2004}. Technological potential of~MPc is impressive and spans over a variety of applications, such as chemical sensors \cite{22Angione2011}, intrinsic semiconductors \cite{21Papageorgiou2004}, field-effect transistors \cite{23Bao1996}, organic light emitting diodes (OLEDs) \cite{24Nguyen2011}, single-molecule devices \cite{25Wu2004}, photovoltaic cells \cite{26Peumans2001}, and materials for nonlinear optics or laser recording \cite{24Nguyen2011}. For this reason, MPc became a model system for nanotechnology and surface chemistry \cite{27Lu1996,28Nazin2003,29Zhao2005,30Zhao2008,31Chen2007}. An isolated Pc molecule (H$_2$Pc) is a planar, aromatic compound built up from four isoindole fragments stuck together by aza-nitrogen atoms in \emph{meso} positions, and C$_{32}$H$_{18}$N$_8$ stoichiometry. MPcs are metallorganic semiconductors with the energy gap usually in the visible spectrum range. The occupation of their highest occupied molecular orbital (HOMO) and lowest unoccupied molecular orbital (LUMO) can be tuned by doping. The most common MPc molecules include Cu, Fe, Mg, Ni, Co, Zn, Ru, Rh, Pd, Cd, and Pt atoms. Their synthesis is quite complex because it implies the formation of the ligand around the metallic center \cite{32Villemin2001}. Amongst MPcs, the FePc, and CoPc play an important role by virtue of magnetic transition metal center, whose electron density of \emph{d}-states near the Fermi level has been detected by high-resolution photoelectron spectroscopy \cite{33Gargiani2010}. Depending on the orientation of the MPc, whether free-standing on a given substrate \cite{34Konig2009} or flat-lying on the Au(110) surface, various number of occupying electrons and different type of empty orbital filling were observed \cite{35Gargiani2010a}. Intimate states originating from the molecule-surface coupling were also observed for CoPc adsorbed on the Au(111) surface \cite{36Hu2008}. %

    To sum up, a proper selection of metal centers in MPc and surface may allow one to control the electronic structure, optical, and magnetic properties of such hybrid systems \cite{37Nardi2013}. Therefore, a comprehensive understanding of their electronic properties using computational quantum chemistry and Density Functional Theory (DFT) is of paramount importance. %

    Here, we use Kohn-Sham hybrid DFT method combined with group theory, and ligand field theory to investigate adsorption, stability, electronic structure, and magnetic properties of graphene - iron(II) phthalocyanine hybrid systems. Standard GGA funtional is applied to probe the PES of FePc adsorption on pristine and defective graphene sheet to determine the ground state structures of hybrid systems for further analysis. We provide also theoretical guidance that may initiate the effort towards synthesis of robust G-MPc hybrid systems for spintronics applications. %
    \section{Methodology}%
    Here, we carry out the plane-wave first-principle calculations in the framework of spin-polarized DFT as implemented in the Vienna Ab-initio Simulation Package (VASP) \cite{38Kresse}. The generalized gradient approximation (GGA) with the Perdew-Burke-Ernzerhof (PBE) \cite{39PerdewBurke} exchange-correlation functional is applied to treat the exchange and correlation energy. Core-valence interactions are treated using projector-augmented wave (PAW) formalism \cite{40PAW}. Valence wavefunctions are expanded into a linear combination of plane-waves with an energy cutoff of 520 eV. Gaussian smearing width of 0.05 eV/\AA~is used to improve convergence of states near the Fermi level. The electronic structures of pristine H$_2$Pc, and FePc, and the most stable G-FePc hybrid systems are determined with hybrid PBE0 functional \cite{41Perdew1996}. The PBE0 functional mixes the PBE exchange energy and Hartree-Fock exchange energy in a set 3 to 1 ratio, along with the full PBE correlation energy: %
    \begin{equation}%
    E_{xc}^{PBE0}=\frac{1}{4}E^{HF}_{x}+\frac{3}{4}E^{PBE}_{x}+E^{PBE}_{c},
    \label{eq:pbe0}
    \end{equation}
    where $E^{HF}_{x}$ is the Hartree-Fock exact exchange functional, $E^{PBE}_{x}$ and $E^{PBE}_{c}$ are the PBE exchange and correlation functionals, respectively. %

    The graphene sheet is modeled by a $22.2\times{22.2}\times{30.0} $ \AA~supercell containing 162 carbon atoms, separated by a vacuum space of 30 \AA. Subsequently, the relaxed graphene is loaded with FePc molecule in a variety of initial configurations. Both, the pristine and the defects-containing graphene sheets have been taken into account. Gas-phase structures of H$_2$Pc and FePc molecules have been calculated in a simulation cubic-box ($35.0\times{35.0}\times{35.0}$ \AA ), and applying only $\Gamma$-point for $k$-point sampling. Various G-FePc hybrid systems have been relaxed until the Hellmann-Feynman forces acting on atoms dropp below 0.01 eV/\AA. In the case of G-FePc structures, we sample the first Brillouin zone using the $\Gamma$-centered $2\times{2}\times{1}$ $k$-point grid. %

    To analyze MAE, we use the approach similar to that one proposed earlier by Wang \emph{et~al.} \cite{42Wang2019}. According to definition, the MAE can be expressed as a difference of the total energy of the system in its different spin orientations, which, in turn, is directly linked with the zero-field splitting (ZFS) parameters $D$ and $E$. In order to calculate MAE, we first carry out a self-consistent, noncollinear calculations without spin-orbit coupling (SOC) included. Using the ground state charge density of the system as an input, the SOC can be treated as a perturbation in non-self-consistent calculations of three perpendicular magnetization directions ($x, y, z$). Values of the longitudinal anisotropy $D$ and the transversal anisotropy $E$ constants can be calculated in the following manner \cite{43Choi2016}: %

    \begin{equation}%
    D^{DFT}=\varepsilon_z-\frac{\varepsilon_x+\varepsilon_y}{2},
    \label{eq:Ddft}
    \end{equation}%
    \begin{equation}%
    E^{DFT}=\varepsilon_x-\varepsilon_y,
    \label{eq:Edft}
    \end{equation}%
    where $\varepsilon_i$ ($i=x,y,z$) denotes the energy when the spin $S$ is aligned with the easy or hard axis $z$ and the two transversal axes $x$ and $y$. %

    \section{Results and discussion}%
    \subsection{Adsorption of iron(II) phthalocyanine on graphene}%
    \begin{figure*}[t]%
    \centering%
    \begin{minipage}{.24\textwidth}%
	   \centering%
	  a)\includegraphics[width=3.5cm]{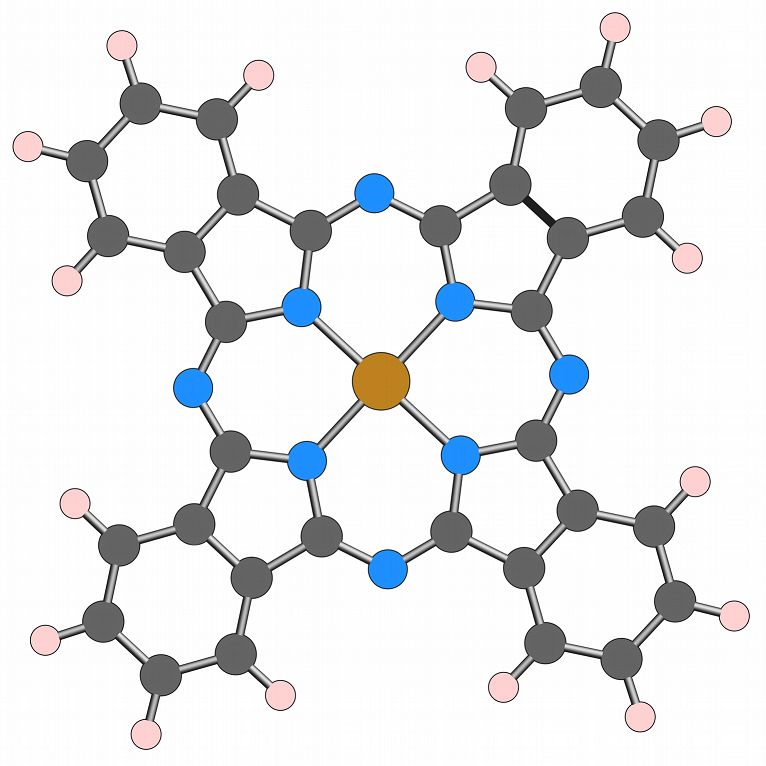}%
	\end{minipage}%
	\begin{minipage}{.74\textwidth}%
	  	\centering%
	  c)\includegraphics[width=10.78cm]{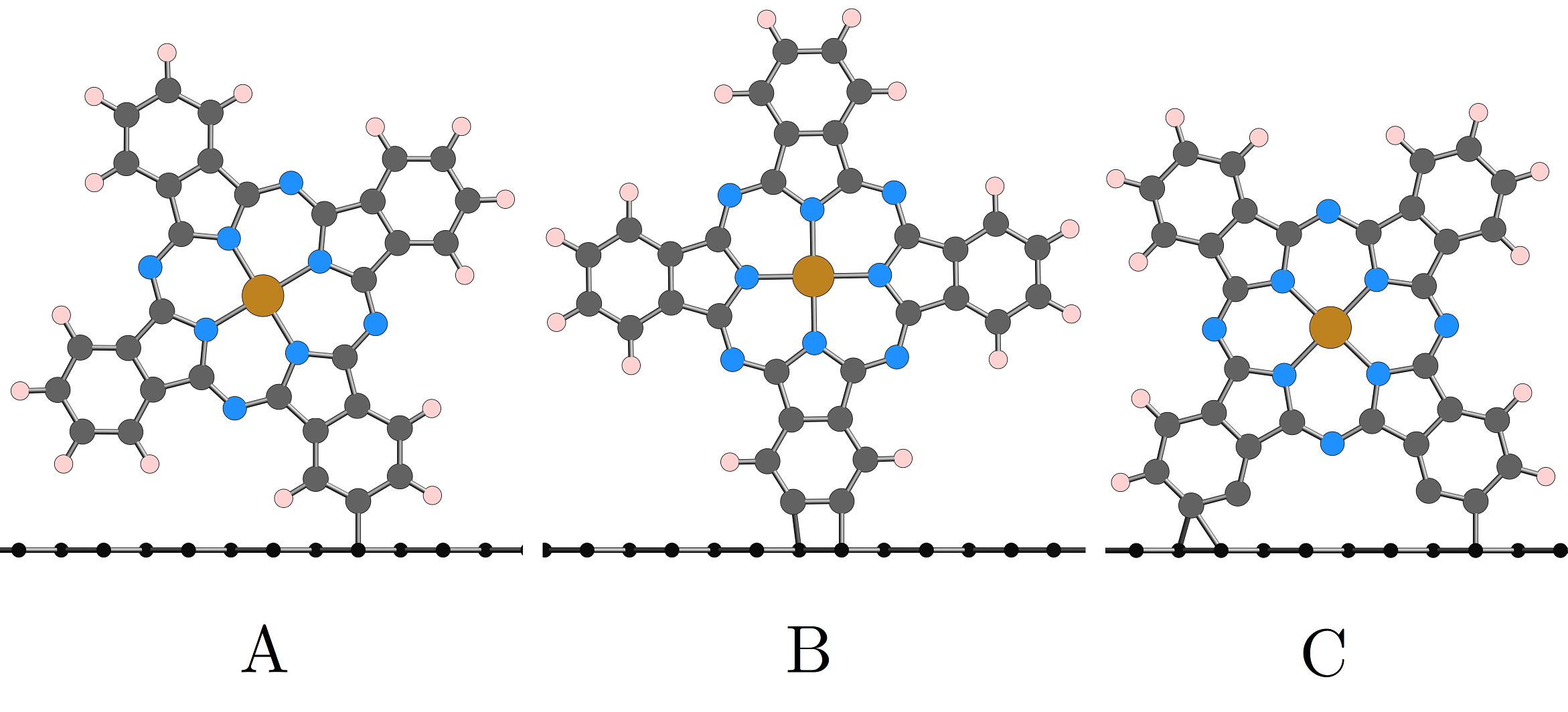}%
	\end{minipage}%
	\newline
	\begin{minipage}{.49\textwidth}%
	  	\centering%
	  b)\includegraphics[width=6.77cm]{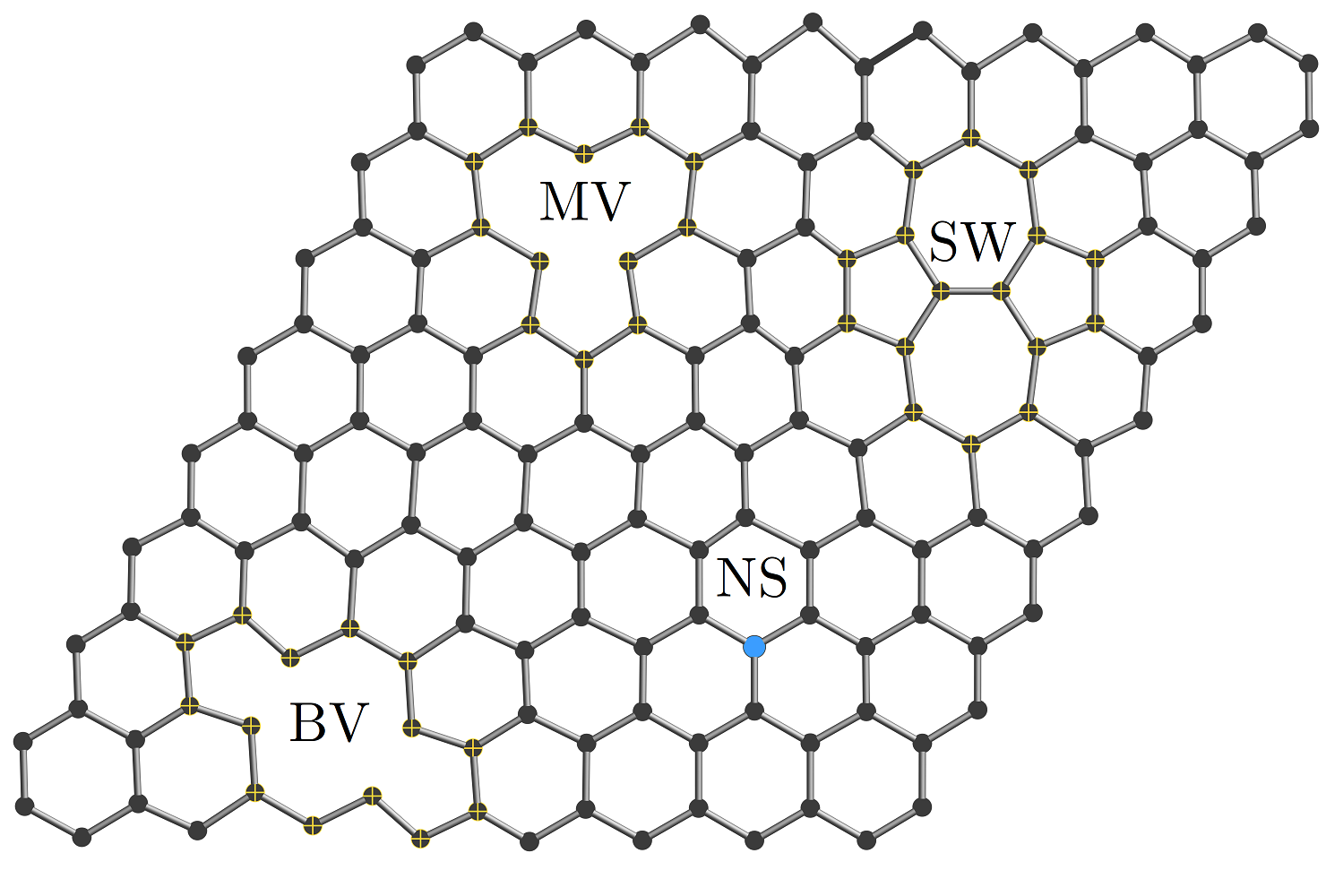}%
	\end{minipage}%
	\begin{minipage}{.49\textwidth}%
	  	\centering%
	  d)\includegraphics[width=6.77cm]{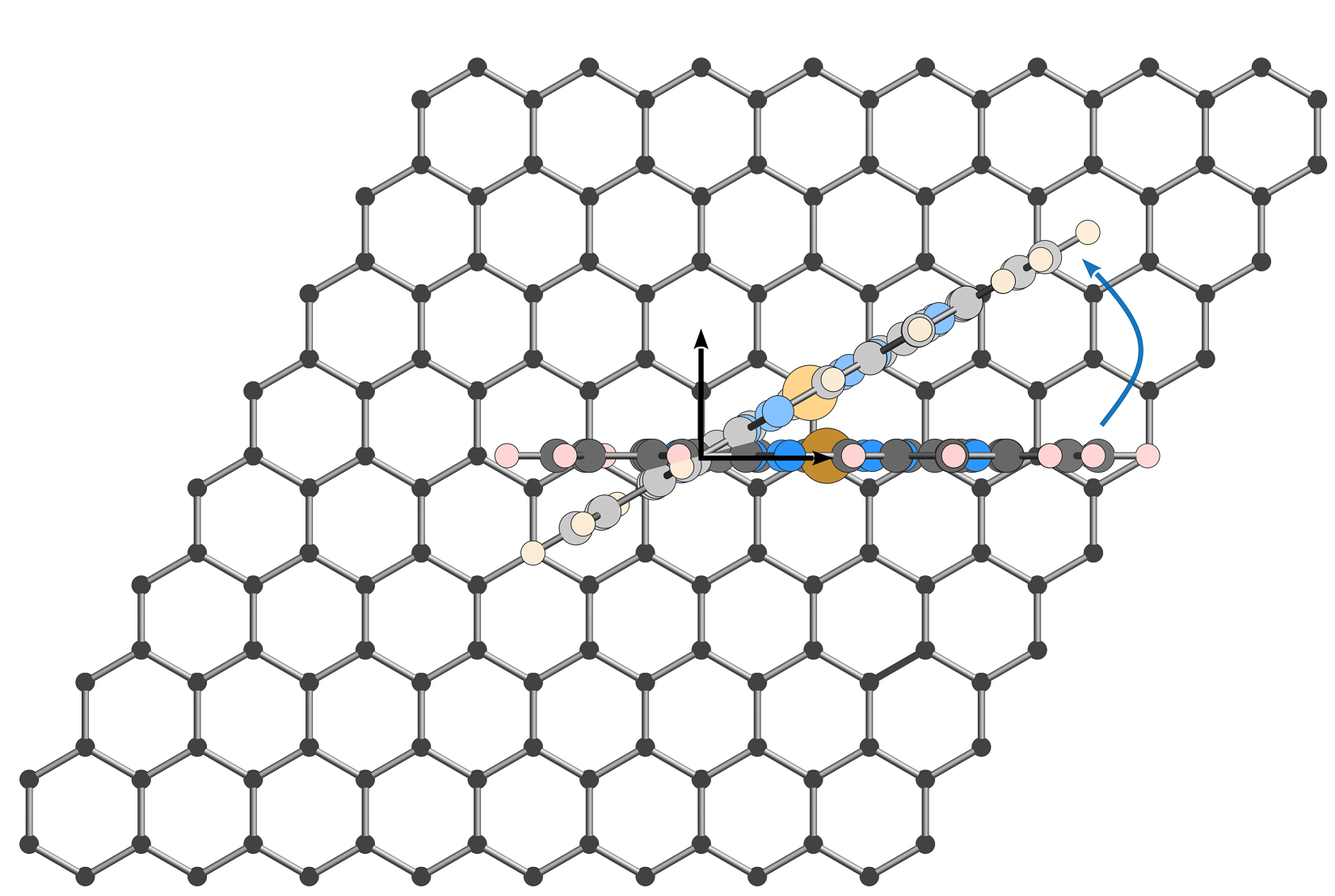}%
	\end{minipage}%
	\newline
	\begin{minipage}{0.33\textwidth}%
	\includegraphics[width=2.487cm]{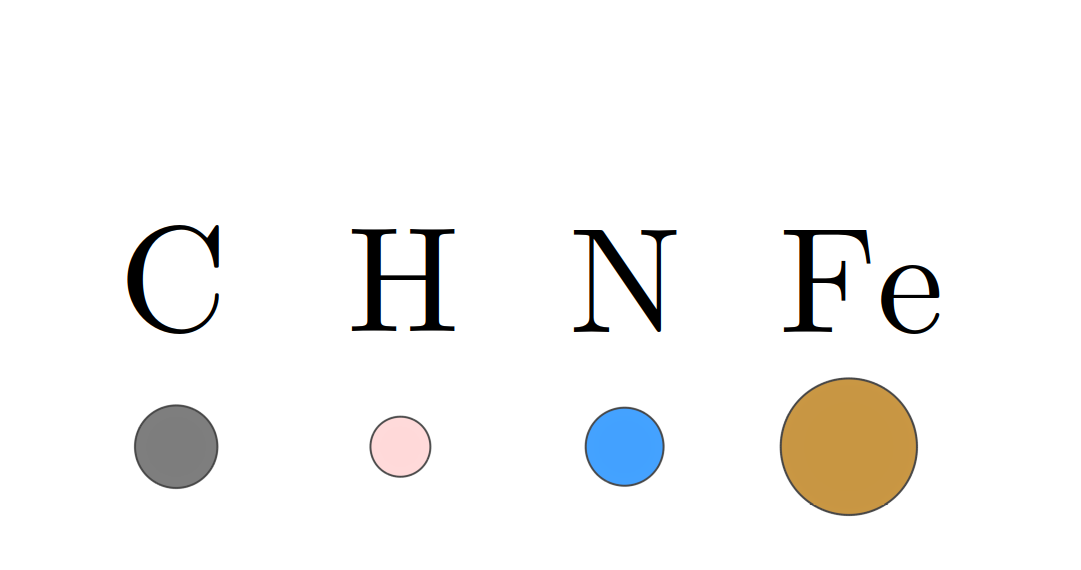}%
	\end{minipage}%
    \caption{Molecular structure of a) iron(II) phthalocyanine; b) graphene sheet with the selected defects indicated (NS $=$ substitional N, SW $=$ Stone-Wales defect, MV $=$ Vacancy, and BV $=$ bivacancy); c) the G-FePc with one (moiety A), two (moiety B) and four (moiety C) hydrogen atoms removed; d) the G-FePc and rotation range on the axis perpendicular to the graphene sheet.
        \label{fig:Fig1}}
    \end{figure*}%

    An individual FePc molecule has a planar structure and belongs to high symmetry $D_{4h}$ point group. The Fe atom occupies the inversion center and it is directly surrounded by four nitrogen neighbors. The nitrogen atoms, in turn, are linked with four aromatic rings, as shown in Fig.\ref{fig:Fig1}a. By removing at least one hydrogen atom from the benzol ring of FePc the carbon dangling bonds are generated. These dangling bonds could hypothetically interact with graphene sheet, enabling one to create an array of perpendicularly stuck FePc on graphene. To prove this hypothesis, we performed large-scale DFT calculations of the adsorption energies. We considered three different FePc moieties with one (A), two (B) and four (C) hydrogen atoms removed (see Fig.\ref{fig:Fig1}c). To sample the potential energy surface (PES) for FePc adsorption on graphene and identify the ground state structures, we generate a number of initial configurations for each individual case by rotating (by angles in range of 0\degree~to 30\degree~by step of 5\degree) the FePc molecule about the axis perpendicular to the graphene sheet or moving the molecule to the different locations on the graphene (see Fig.\ref{fig:Fig1}d). The bridge, onsite (top site), and hollows (H$_6$ and H$_3$) adsorption sites on the graphene are considered. In addition, we investigate the impact of point defects in graphene, \emph{i.e.} monovacancy (MV), bivacancy (BV), substitutional nitrogen (SN) and Stone-Wales (SW) (see Fig.\ref{fig:Fig1}b), on the adsorption energy of FePc. Even though we end up with almost 80 initial configurations most of them reconfigured and converged to several new structures that turned out to be the ground states. The calculated adsorption energies are juxtaposed in Table \ref{tab:Adsorption}. 
\begin{table}[h]%
\centering%
\caption{Adsorption energy of FePc on pristine (PG) and defective graphene (NS $=$ substitional N, SW $=$ Stone-Wales defect, MV $=$ Vacancy, and BV $=$ bivacancy) sheet for different types of adsorption (moieties A, B, and C) calculated with PBE functional. Both, the dissociative adsorption with spontaneous elimination of hydrogen ($E^{(1)}_{ad}$) and adsorption of activated FePc ($E^{(2)}_{ad}$) are presented (see eqs \ref{eq:Ead1} and \ref{eq:Ead2}, respectively, in the main text). } 
\begin{tabular}{cccc}%
Base                & Type & $E^{(1)}_{ad}$ [eV]    & \multicolumn{1}{c}{$E^{(2)}_{ad}$ [eV]}    \\ \hline
\multirow{3}{*}{PG} & A    & 1.58  & \multicolumn{1}{c}{-0.96} \\
                    & B    & 3.29  & \multicolumn{1}{c}{-2.22} \\
                    & C    & 7.71  & \multicolumn{1}{c}{-0.92} \\ \hline
MV                  & A    & -0.46 & -3.00                     \\ \hline
\multirow{2}{*}{BV} & A    & 0.87  & -1.67                     \\
                    & B    & -3.20 & -8.71                     \\ \hline
\multirow{2}{*}{NS} & A    & 12.75 & 10.21                     \\
                    & B    & 8.52  & 3.01                      \\ \hline
SW                  & A    & 2.09  & -0.44                     \\ \hline
\end{tabular}
\label{tab:Adsorption}%
\end{table}%
We consider two different possibilities of adsorption process: (i) dissociative adsorption with spontaneous elimination of hydrogen, and (ii) adsorption of activated FePc. For these two cases, the adsorption energy can be extracted from the following formulas:  
    \begin{equation}
        E^{(1)}_{ad}=E_{G-FePc}-E_G-E_{FePc}+n\mu_{H},
    \label{eq:Ead1}
    \end{equation}
    \begin{equation}
        E^{(2)}_{ad}=E_{G-FePc}-E_G-E^{'}_{FePc},
    \label{eq:Ead2}
    \end{equation}
where $E_{G-FePc}$ denotes the total energy of graphene-FePc hybrid system, $E_G$ denotes the total energy of graphene sheet (either pristine or defective), $n$ is the number of H atoms removed, $\mu_{H}$ denotes the chemical potential of hydrogen which here is equal to half of the total energy of H$_2$ molecule, and $E^{'}_{FePc}$ denotes the total energy of the activated FePc molecule in the gas phase with one, two or four hydrogen atoms removed. As can be noticed in Table \ref{tab:Adsorption}, the dissociative adsorption of FePc on graphene is thermodynamically unfavorable due to positive value of adsorption energy. The adsorption energy displays linear behavior with a number of hydrogen atoms removed. The situation changes substantially when the adsorption of activated FePc is considered. In this case, the adsorption energy is negative, making the process thermodynamically favorable. In the next step, we calculated the adsorption energy of FePc on the defective graphene sheet. Generally, the presence of point defects additionally enhances the perpendicular adsorption of FePc on graphene by shifting the adsorption energy towards more negative values. %

To sum up, we demonstrated that perpendicularly stuck FePc on graphene sheet may be indeed stable and the synthesis of G-FePc hybrid systems possible, assuming proper activation of FePc before deposition on graphene. Point defects engineering in the graphene is beneficial and may be helpful to achieve this type of structure. %
\subsection{Electronic structure of G-FePc hybrid systems}%

\begin{figure*}[t]
\centering%
    \begin{minipage}{.33\textwidth}%
	 \centering%
	 a)\includegraphics[width=3.592cm]{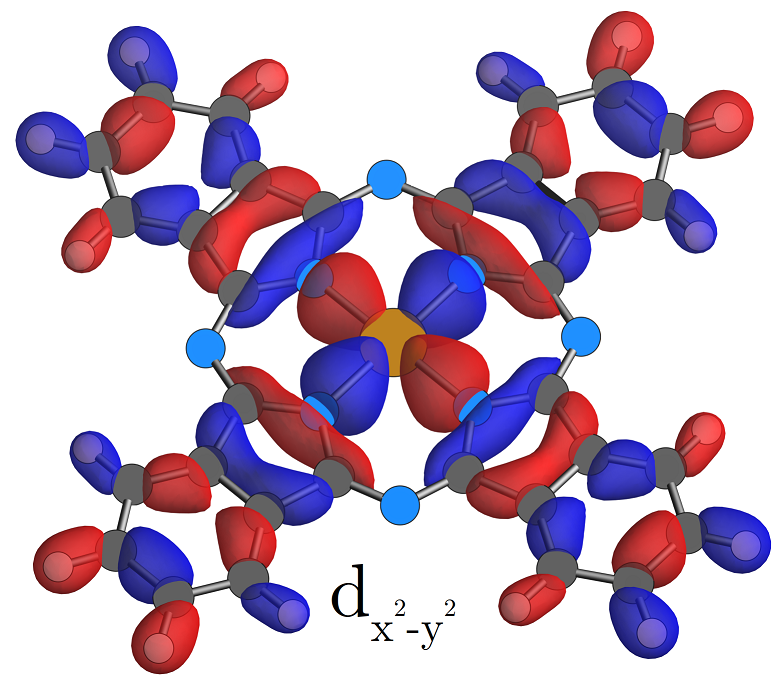}%
	\end{minipage}%
	\begin{minipage}{.33\textwidth}%
	 \centering%
	 b)\includegraphics[width=3.466cm]{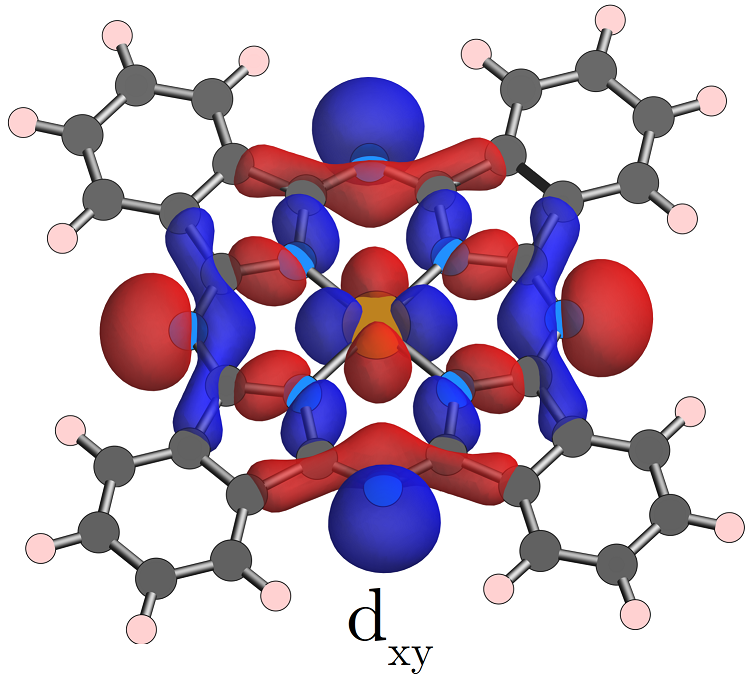}%
	\end{minipage}%
	\begin{minipage}{0.33\textwidth}%
	 \centering%
	 c)\includegraphics[width=3.456cm]{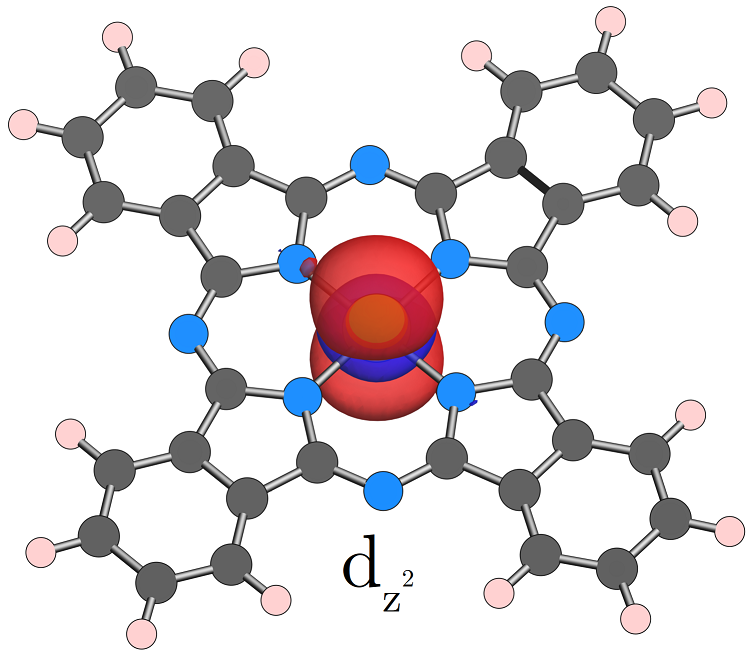}%
	\end{minipage}%
	\newline
	\begin{minipage}{0.33\textwidth}%
	 \centering%
	 d)\includegraphics[width=3.456cm]{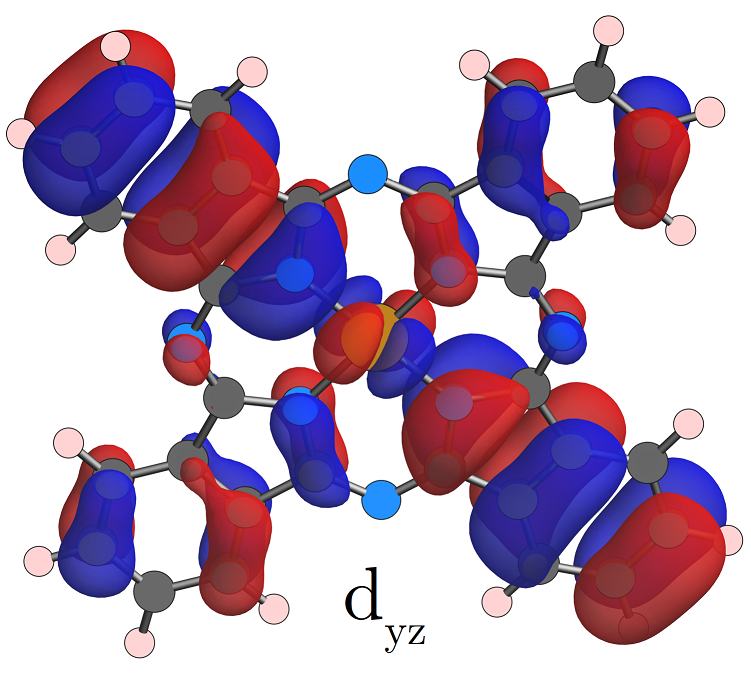}%
	\end{minipage}%
	\begin{minipage}{0.33\textwidth}%
	 \centering%
	 e)\includegraphics[width=3.526cm]{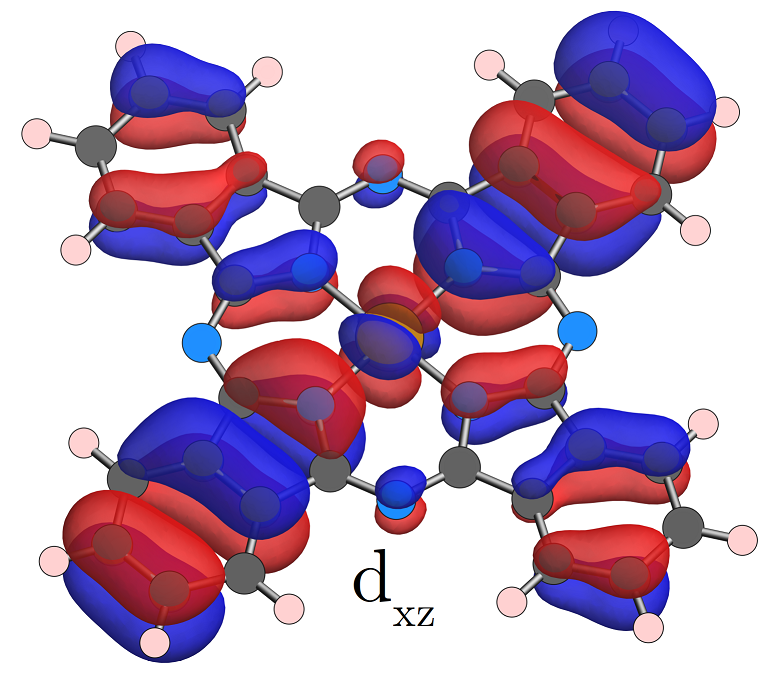}%
	\end{minipage}%
	\begin{minipage}{0.33\textwidth}%
	 \centering%
	 f)\includegraphics[width=3.526cm]{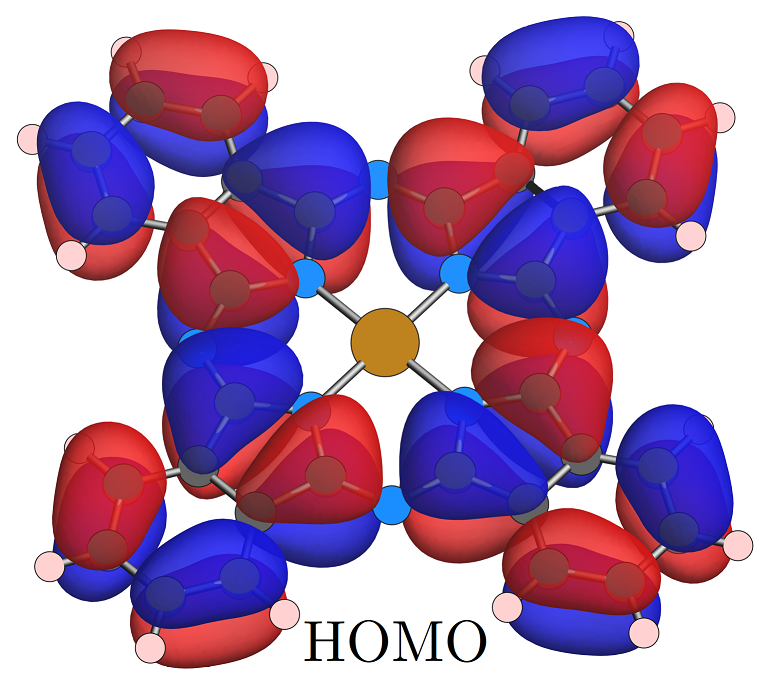}%
	\end{minipage}%
	\newline
	\begin{minipage}{0.33\textwidth}%
	 \centering%
	 g)\includegraphics[width=3.526cm]{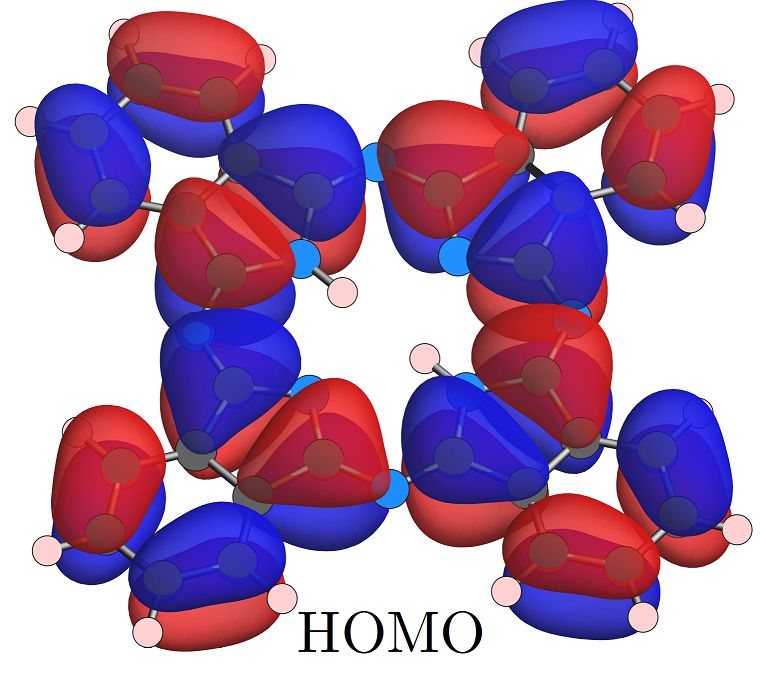}%
	\end{minipage}%
	\begin{minipage}{0.33\textwidth}%
	 \centering%
	 h)\includegraphics[width=3.592cm]{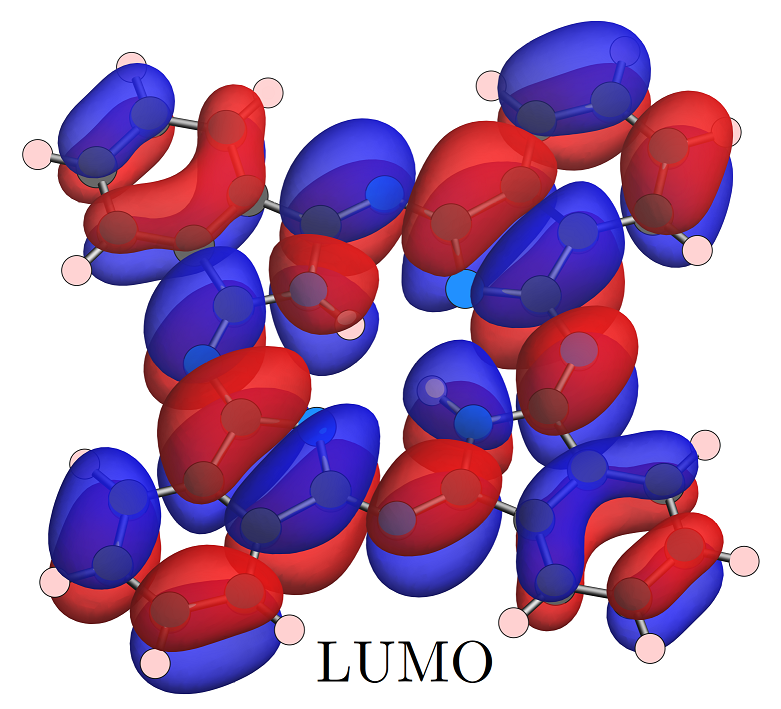}%
	\end{minipage}%
	\begin{minipage}{0.33\textwidth}%
	 \centering%
	\includegraphics[width=2.487cm]{Fig1.png}%
	\end{minipage}%
\caption{Visualization of the selected Kohn-Sham wavefunctions of the FePc molecule. The orbitals are extracted from the PBE0 hybrid calculations using homemade code \cite{57CzelejTi,58Czelej2018a,59Czelej2018}. The red (blue) lobes indicate the positive (negative) phase of the wavefunctions with an arbitrarily selected isosurface value. a) $d_{x^2-y^2}$, b) $d_{xy}$, c) $d_{z^2}$, d) $d_{yz}$, e) $d_{xz}$ orbitals, f) HOMO of FePc, g) HOMO of H$_2$Pc, and h) LUMO of H$_2$Pc.
        \label{fig:Orbitals}}
\end{figure*}%

\begin{figure*}[t]%
\centering%
\includegraphics[width=9.36cm]{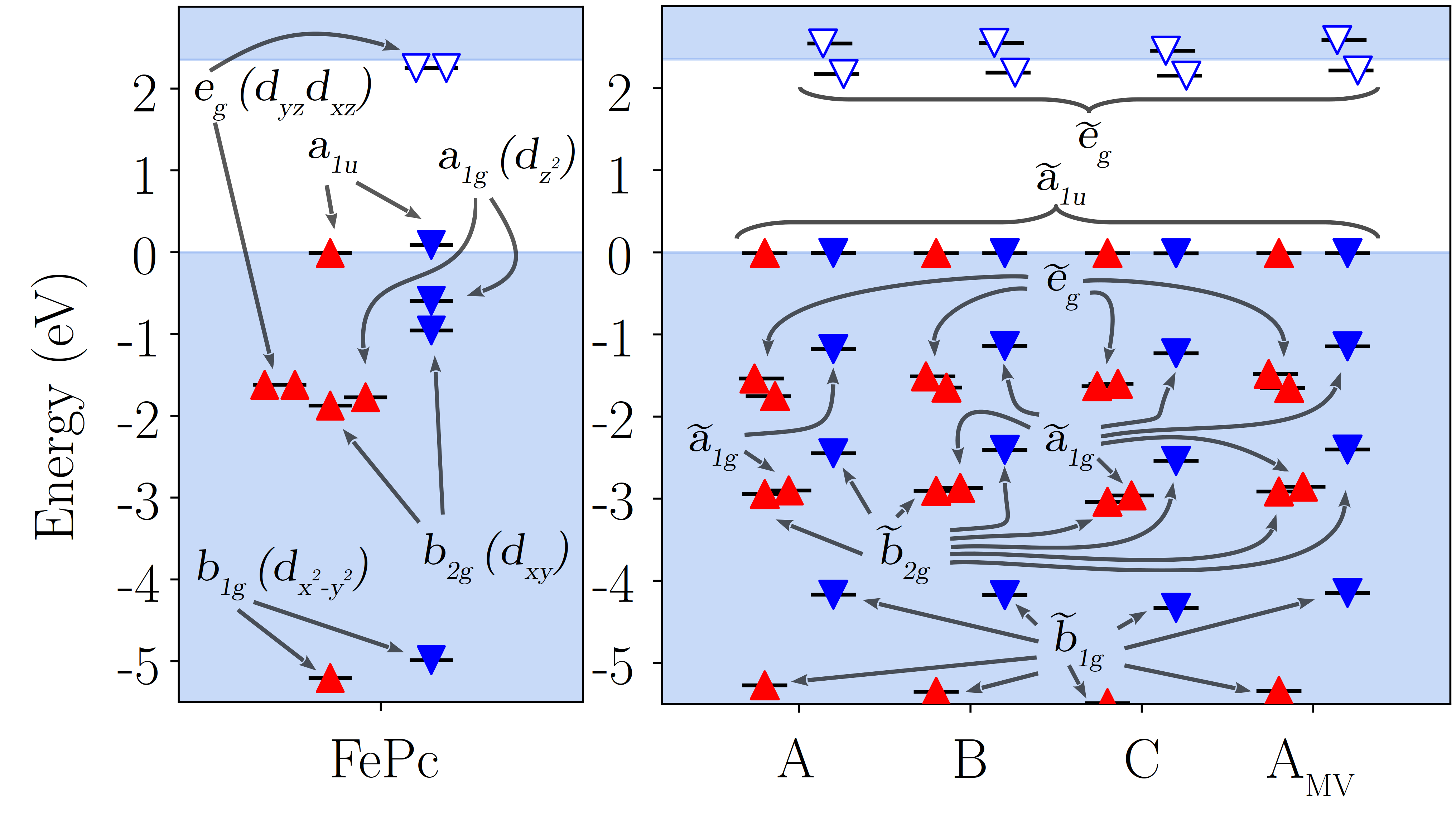}%
\caption{Kohn-Sham eigenvalue spectra of FePc molecule and G-FePc hybrid structures calculated with the PBE0 hybrid functional. The white area represents HOMO - LUMO gap of single H$_2$Pc. Spin-up (-down) channels are indicated by red (blue) triangles, whereas the filled (unfilled) triangles represent the occupied (empty) states. For the sake of clarity, only the $3d$-related states, HOMO, and LUMO are presented in the graph. In the case of G-FePc, we label the electronic states with tilde, in order to indicate their origin with respect to single FePc.
        \label{fig:KSlevel}}
\end{figure*}

Prior to the analysis of the electronic structure of G-FePc hybrid systems, it is worth to reinvestigate the electronic structure of pristine FePc molecule in details. As we have already mentioned, the FePc molecule has a $D_{4h}$ symmetry and exhibits paramagnetic $S=1$ spin state, confirmed by our PBE0 hybrid calculations, previous high-level \emph{ab initio} calculations\cite{42Wang2019} and experiments \cite{44Bidermane2015,45Kroll2012}. In $D_{4h}$ crystal field generated by four nitrogen ligands, the $3d$ orbitals of~Fe split into non-degenerate $d_{z^2}$, $d_{x^2-y^2}$, $d_{xy}$ and doubly-degenerate $d_{xz}$ and $d_{yz}$. According to the $D_{4h}$ character table, these states transform in the molecule as follows: $d_{z^2}\rightarrow{a_{1g}}$, $d_{x^2-y^2}\rightarrow{b_{1g}}$, $d_{xy}\rightarrow{b_{2g}}$ and $(d_{xz}, d_{yz})\rightarrow{e_g}$. The 3D contour plot of each relevant orbital is presented in Fig.\ref{fig:Orbitals}. As the lobes of $d_{x^2-y^2}$ orbital are pointed directly toward the nearest nitrogen ligands one can expect the highest overlap with the $s$-$p$ hybrid states. This strong interaction results in the formation of bonding and anti-bonding combination of $b_{1g}$ states significantly split in energy; the former resides deep beneath HOMO and the latter high above LUMO of FePc. For the sake of clarity, only the bonding $b_{1g}$ is shown in Fig.\ref{fig:Orbitals}. The remaining orbitals $d_{z^2}$, $d_{xy}$, $d_{xz}$ and $d_{yz}$ exhibit only minor overlap with nitrogen $s$-$p$ states; therefore, the $a_{1g}$, $b_{2g}$ and $e_g$ should be relatively close in energy. Our PBE0 hybrid calculations confirm this statement (see Fig.\ref{fig:KSlevel}). In contrast to all of the Fe-related states being symmetric with respect to inversion center, the $a_{1u}$ assigned as HOMO of FePc is anti-symmetric (see Fig.\ref{fig:Orbitals}f) and hence, does not mix with $3d$ orbitals at all. %

    The ground state electronic configuration of FePc corresponding to $S = 1$ spin state has been recently a subject of intense debate \cite{33Gargiani2010,42Wang2019,44Bidermane2015,45Kroll2012,46Liao2005}. Amongst four different possible configurations for the triplet spin, two electronic states, \emph{i.e.} $^3$A$_{2g}$ and $^3$E$_g$ corresponding to $b_{1g}^{(2)} b_{2g}^{(2)} e_g^{(2)} a_{1g}^{(2)} b_{1g}^{(0)}$ and $b_{1g}^{(2)} b_{2g}^{(2)} e_g^{(3)} a_{1g}^{(1)} b_{1g}^{(0)}$ molecular orbital occupations are assigned to the ground state. Unambiguous assignment of FePc ground state turned out to be very challenging due to small energy separation between $^3$A$_{2g}$ and $^3$E$_g$ electronic states \cite{45Kroll2012,46Liao2005} and possible multi-determinant nature induced by the spin-orbit interaction \cite{47Fernandez-Rodriguez2015}. It has to be mentioned that gas phase experiments of MPc molecules require high temperature to create a constant molecular vapor flux and at the same time, keeping the experimental set-up reliable and under control. Under such conditions, along with the spectral line broadening, thermal excitation between two closely located electronic states may occur and vibration-induced mixing can be even more pronounced. As can be seen in Fig.\ref{fig:KSlevel}, our PBE0 hybrid DFT calculations yield the $^3$A$_{2g}$ electronic ground state of FePc. Assuming low temperature case, this assignment seems reasonable as the $b_{1g}^{(2)} b_{2g}^{(2)} e_g^{(2)} a_{1g}^{(2)} b_{1g}^{(0)}$ occupation of the $^3$A$_{2g}$ electronic state represents the Jahn-Teller stable system. By contrast, an extra electron located at spin-minority $e_g$ state in case of $^3$E$_g$ configuration causes Jahn-Teller instability that would lift the degeneracy of $e_g$ and ultimately, break the $D_{4d}$ symmetry of molecular orbitals. Based on the 3D plots of the orbitals as presented in Figs.\ref{fig:Orbitals}f and \ref{fig:Orbitals}g, the HOMO of FePc and H$_2$Pc have exactly the same nature and symmetry, whereas, the LUMO of FePc is a doubly-degenerate $e_g$ state. The calculated HOMO - LUMO gap of 2.16 eV is located at the spin-minority channel, which is consistent with sharp Q-band at 1.92 eV in the experimentally observed absorption spectra \cite{48Sumimoto2009}. %

    Having the electronic structure of a single FePc established, we focused our attention on the G-FePc hybrid systems. It is commonly known that upon adsorption the electronic structure may significantly change due to symmetry lowering, formation of new bonds between substrate and molecule, or charge transfer effects. %

\begin{figure*}[t]%
\centering%
	\begin{minipage}{0.48\textwidth}%
	\centering%
	\includegraphics[width=7.53cm]{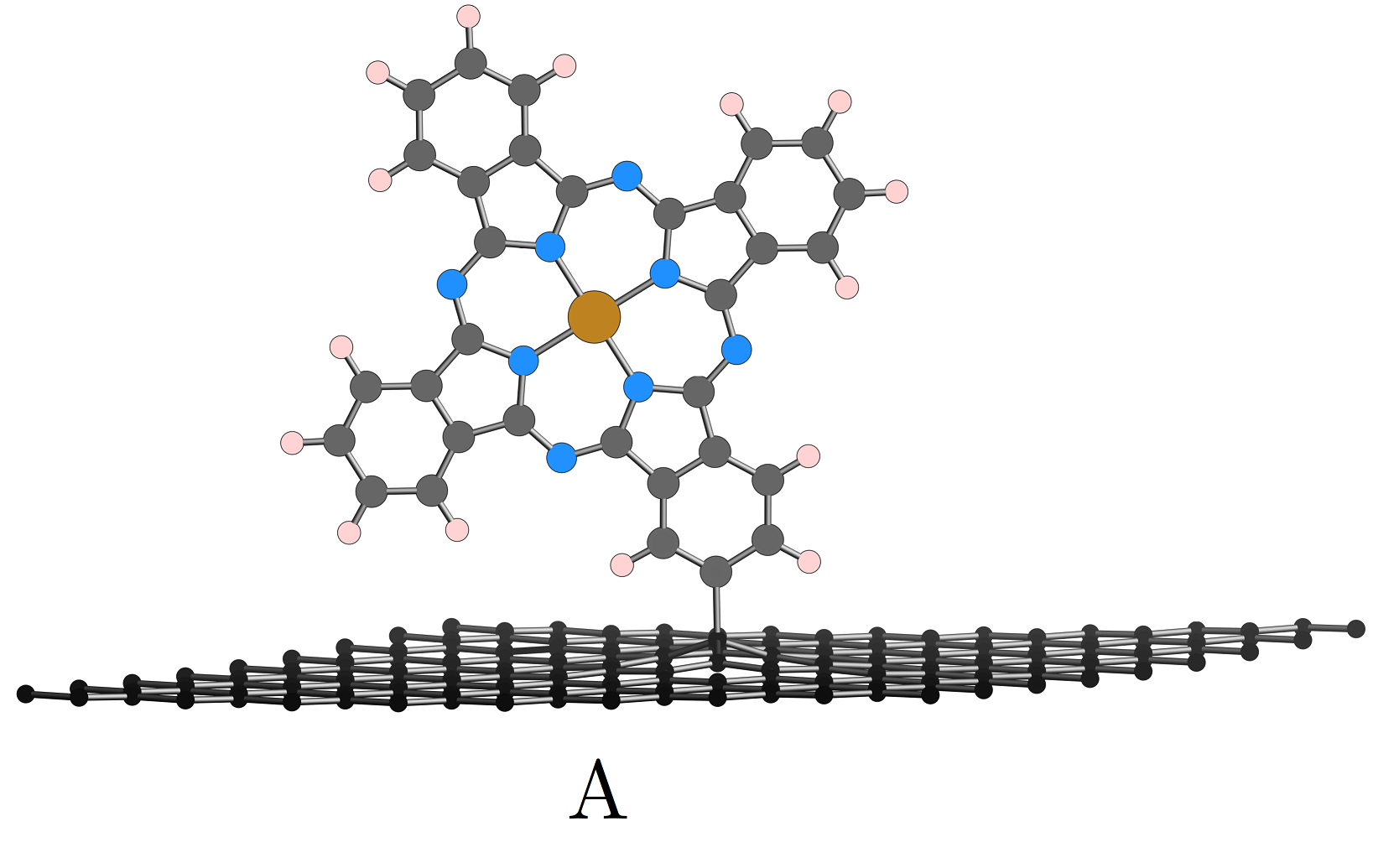}%
	\end{minipage}%
	\begin{minipage}{0.52\textwidth}%
	\centering%
	\includegraphics[width=8.5cm]{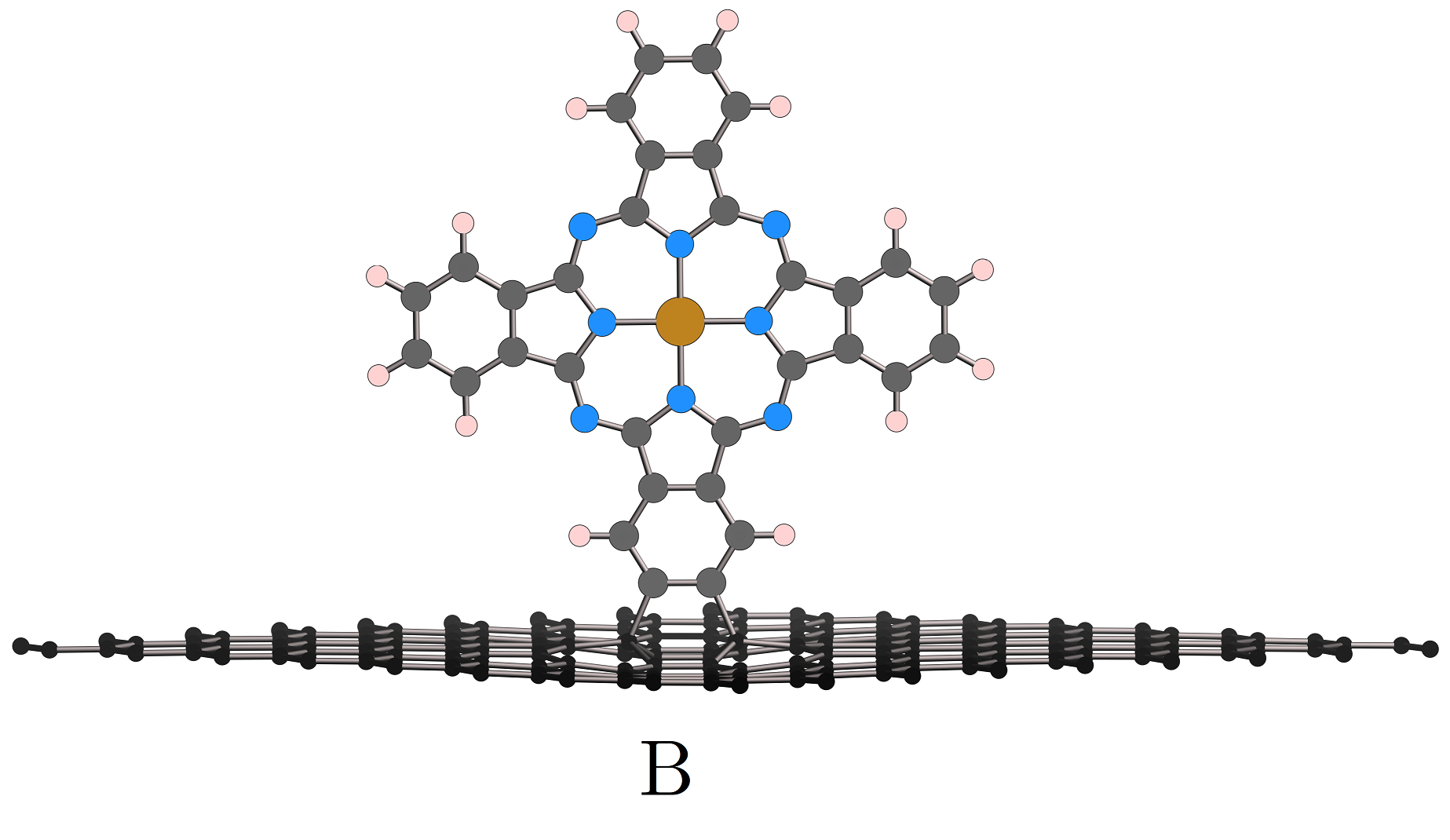}%
	\end{minipage}%
	\newline
	\begin{minipage}{0.52\textwidth}%
	\centering%
	\includegraphics[width=8.5cm]{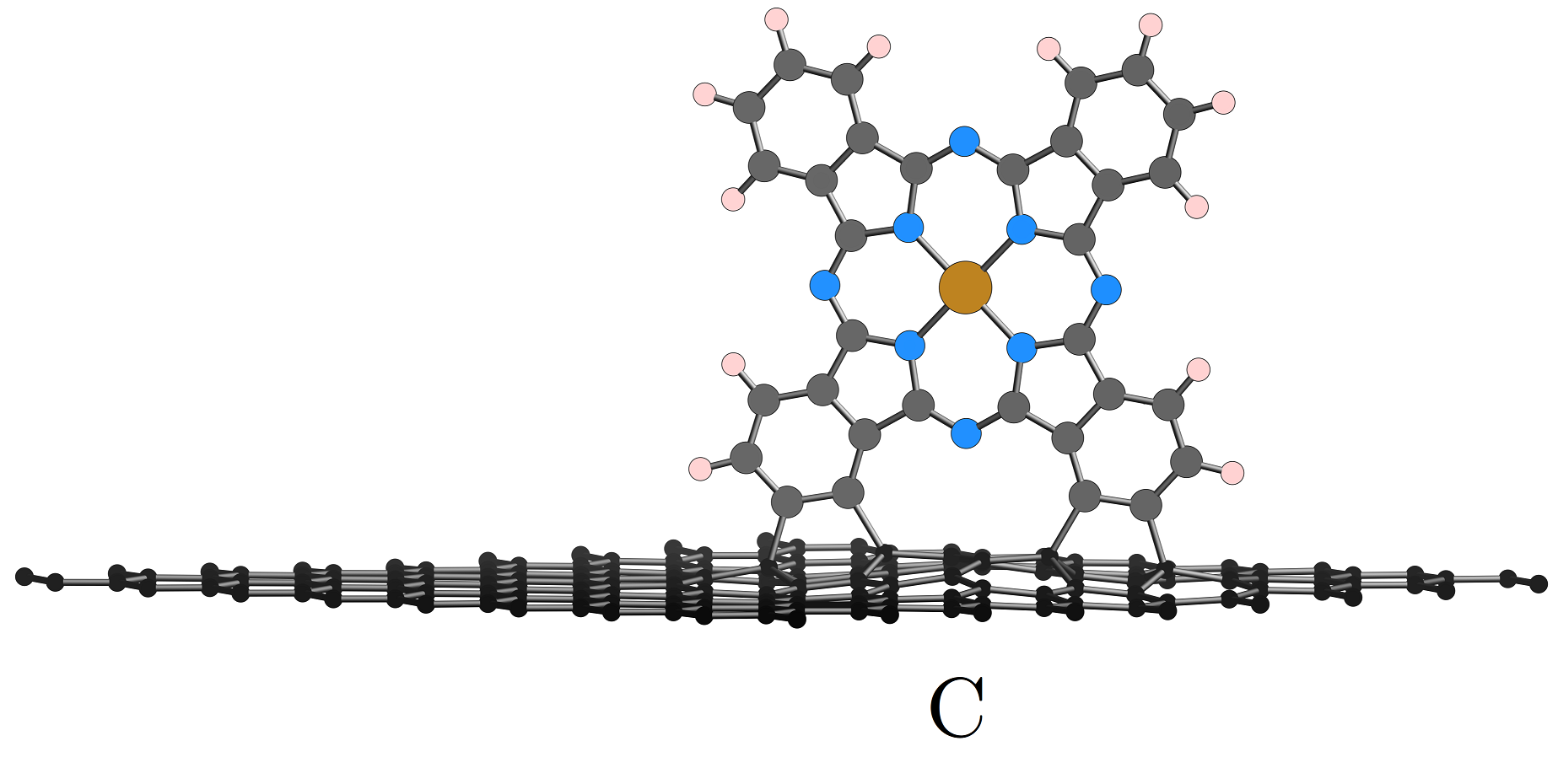}%
	\end{minipage}%
	\begin{minipage}{0.48\textwidth}%
	\centering%
	\includegraphics[width=7.53cm]{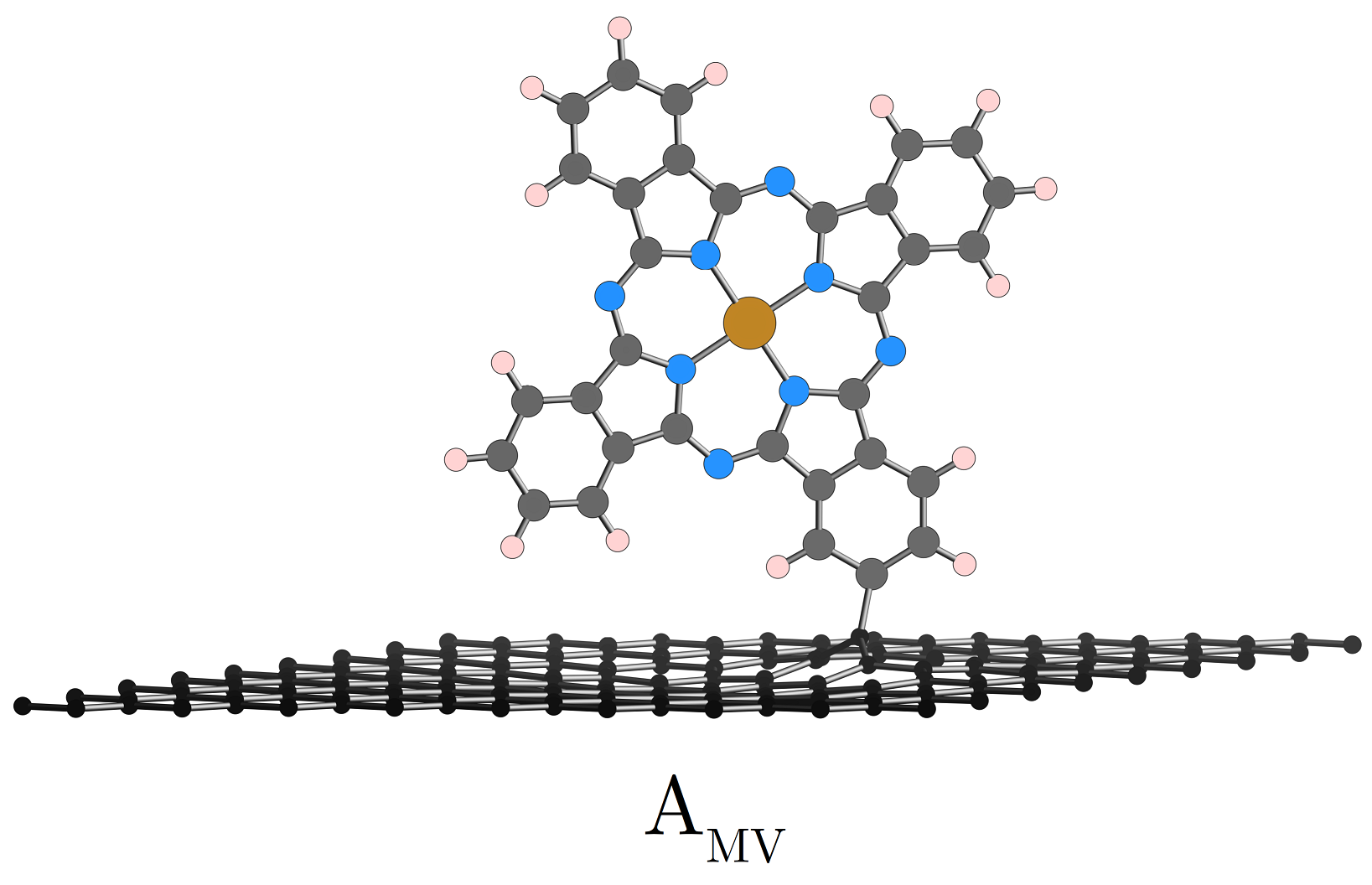}%
	\end{minipage}%
	\newline
	\begin{minipage}{0.33\textwidth}%
	\centering%
	\includegraphics[width=2.487cm]{Fig1.png}%
	\end{minipage}%
\caption{Relaxed structures of FePc adsorbed on pure graphene with one (A), two (B), and four (C) hydrogen atoms removed, and FePc adsorbed on graphene with monovacancy (A$_{MV}$).
        \label{fig:Relaxed}}
\end{figure*}

    First, we notice that regardless of the adsorption conformation, the onsite magnetic moment induced on Fe atom remains unchanged and the moiety exhibits triplet $S = 1$ spin state. To provide a deeper insight into the electronic structure of G-FePc hybrid systems and explain the stability of $S = 1$ spin state, we selecte four representative conformations and carry out structural relaxation using PBE0 hybrid functional. Three of the selected structures are the ground states corresponding to perpendicular adsorption of FePc on pure graphene with one (A), two (B), and four (C) hydrogen atoms removed, and one structure is the ground state of FePc molecule with one hydrogen removed and adsorbed on monovacancy-containing graphene (A$_{MV}$; see Fig.\ref{fig:Relaxed}). The PBE0 eigenvalue spectrum of these four structures is presented in Fig.\ref{fig:KSlevel}. For the sake of clarity, we depict the $3d$-related states, HOMO and LUMO, in order to find out how these states transform upon adsorption. %

    We can clearly notice that perpendicular adsorption neither change the orbital energy order nor the electron occupancy. The doubly-degenerate $e_g$ state, however, gets slightly split as the symmetry of entire system is lowered. Nevertheless, the local symmetry in close proximity to Fe$^{2+}$ ion is nearly preserved, keeping the $3d$-related orbitals very similar to those in an isolated FePc. Thus, we can conclude that magnetic properties of perpendicularly adsorbed FePc on graphene should not be affected by FePc - graphene interactions. %

    \subsection{Magnetic properties of G-FePc hybrid systems}%
    The interaction of electron spin on magnetic ion with the surrounding crystal field can be described by the following Hamiltonian: %
\begin{equation}%
\hat{H}=D\hat{S}^{2}_z +E(\hat{S}^{2}_x-\hat{S}^{2}_y),
\label{eq:HZFS}
\end{equation}%
    where $\hat{S}=(\hat{S}_x,\hat{S}_y,\hat{S}_z)$ are the spin operators, $D=D_{zz}-\frac{1}{2}(D_{xx}+D_{yy})$ denotes the longitudinal anisotropy strength, and $E=\frac{1}{2}(D_{xx}-D_{yy})$ denotes the transversal anisotropy strength. The $(D_{xx},D_{yy},D_{zz})$ are the three components of the so-called zero-field splitting tensor (ZFS) and they are directly related to the environmental symmetry. The Hamiltonian written above describes the relation between magnetic energy and the orientation of individual spins with respect to their environment, or in the other words, the MAE of the system. In the case of organometallic structures the performance of conventional DFT methods in accurate prediction of the MAE is rather poor \cite{49Haldar2018,50Wang2016,51Hu2013} mainly due to the static correlation error \cite{52Cohen2012}. The discrepancy between DFT prediction and the experiment for G-FePc composites can even reach one order of magnitude \cite{42Wang2019}. %

    First, we focus our attention on the determination of MAE parameters for the isolated FePc molecule and compare our results to the experimentally observed values ($D = 8.7$~meV, $E = 0$~meV) \cite{53Dale1968}. To verify the performance of DFT in the ZFS parameters, we calculate $D$ and $E$, by applying: (i) framework implemented in the VASP code\cite{ZFSVASP} for collinear spin-polarized PBE and PBE0 functionals; (ii) noncollinear PBE$+$SOC (from the eqs \ref{eq:Ddft} and \ref{eq:Edft}), and (iii) noncollinear PBE0 hybrid functional taking into account SOC (from the eqs \ref{eq:Ddft} and \ref{eq:Edft}). The (i) approach results in evident underestimate of ZFS parameters for PBE ($D = 0.13$~meV, $E = 0.00$~meV) as well as PBE0 ($D = 0.29$~meV, $E = 0.00$~meV) functionals. We notice that the application of noncollinear PBE0 hybrid functional with SOC yields $D = 4.24$~meV, $E = 0.00$~meV (for PBE+SOC $D = 4.06$~meV, $E = 0.00$~meV), which is only twice underestimated with respect to experimental value, and to the best of our knowledge, the most accurate estimation of ZFS parameters for single FePc molecule, based on straightforward DFT calculations. Using the noncollinear PBE0 hybrid functional with SOC approach we determine the average SOC energy of Fe atom in the FePc molecule, which is approximately 85\% of SOC energy of free Fe$^{2+}$ ion ($-24.8$~meV)\cite{Tsukahara2016,SOCFe}. %

    The improvement of our results originates from the application of PBE0 hybrid functional that can partially mimic the static correlation. This phenomenon, which is associated with the presence of Fock operator in the PBE0 functional, has been extensively discussed by Yu \emph{et al.} \cite{54Yu2016}. %

    To estimate the MAE parameters for G-FePc hybrid systems, we have applied the well-established procedure that has been successfully applied to similar molecular complexes in previous studies \cite{42Wang2019,55Wang2018a,56Wang2018}. According to this approach, the longitudinal anisotropy strength can be estimated employing the formula: %
\begin{equation}%
D=D^{exact}_{FePc}+(D^{DFT}_{G-FePc}-D^{DFT}_{FePc}),
\label{eq:Dapprox}
\end{equation}%
    where $D^{exact}_{FePc}$ is the exact value of longitudinal MAE for an isolated FePc molecule, whereas, the expression in parenthesis denotes the difference between longitudinal MAE for G-FePc hybrid system and isolated FePc molecule calculated from DFT. The calculated MAE parameters $D$ and $E$ for G-FePc hybrid systems are summarized in Table \ref{tab:ZFSParam}. As can be noticed, the longitudinal anisotropy strength changes slightly, while the transversal anisotropy strength equals nearly zero. Since the MAE parameters give an insight into symmetry of local crystal field near Fe$^{2+}$ ion, the nonzero value of the transversal anisotropy strength would reflect the reduced symmetry of molecule upon the adsorption. Similar to the case of isolated FePc, it is obvious that the local symmetry in close proximity of Fe$^{2+}$ ion is well preserved. %
\begin{table}[h]%
\centering%
\caption{The ZFS parameters $D$ and $E$ for FePc adsorbed on pure graphene with one (A), two (B), and four (C) hydrogen atoms removed, and FePc adsorbed on graphene with monovacancy (A$_{MV}$) calculated (see eq \ref{eq:Dapprox} in the main text) with noncollinear PBE$+$SOC approach.} 
\begin{tabular}{ccc}
Type     & $D$ [meV]   & $E$ [meV]\\ \hline
A        & 8.13 & 0.11  \\
B        & 8.65 & -0.03 \\
C        & 8.96 & 0.04  \\
A$_{MV}$ & 8.90 & 0.04  \\ \hline
\end{tabular}
\label{tab:ZFSParam}%
\end{table}%
    \section{Summary and conclusions}%
    In summary, we have investigated the electronic structure and magnetic properties of G-FePc hybrid systems. Based on the FePc adsorption energy calculations, we demonstrate that perpendicularly stuck FePc on graphene sheet may be indeed stable and the synthesis of G-FePc hybrid systems possible, assuming proper activation of FePc before deposition on graphene. Generally, the presence of point defects on the graphene sheet additionally facilitates the adsorption of FePc, and it is beneficial for achieving high density of SMMs per unit area. Using the combination of group theory, ligand field splitting, and the calculated PBE0 Kohn-Sham eigenvalue spectrum, we demonstrate that the low temperature ground state of isolated FePc molecule is the triplet $^3$A$_{2g}$ and the $S = 1$ spin state is preserved, regardless of adsorption site and the number of removed hydrogen atoms from the benzol rings of FePc. By analyzing the electronic structure and the estimated MAE parameters for G-FePc hybrid systems, we prove that intrinsically high magnetic anisotropy of FePc is only slightly affected by FePc - graphene interaction, opening up new avenue in designing scalable graphene - SMMs systems, suitable for spintronics applications. In addition, we reveal the accuracy limit of single-determinant Kohn-Sham DFT approach in the determination of ZFS parameters for organometallic structures. 
    
	\begin{acknowledgments}
     This research was financially supported by the Polish National Science Centre under contract no. UMO-2016/23/B/ST3/03567. Computing resources were provided by High Performance Computing facilities of the Interdisciplinary Centre for Mathematical and Computational Modeling (ICM) of the University of Warsaw under Grant No. GB69-32. All structures were created and visualised using OVITO and VESTA softwares \cite{60VESTA,61Ovito}. 
	\end{acknowledgments}
	
\bibliography{SMM}%

\begin{thebibliography}{64}%
\makeatletter
\providecommand \@ifxundefined [1]{%
 \@ifx{#1\undefined}
}%
\providecommand \@ifnum [1]{%
 \ifnum #1\expandafter \@firstoftwo
 \else \expandafter \@secondoftwo
 \fi
}%
\providecommand \@ifx [1]{%
 \ifx #1\expandafter \@firstoftwo
 \else \expandafter \@secondoftwo
 \fi
}%
\providecommand \natexlab [1]{#1}%
\providecommand \enquote  [1]{``#1''}%
\providecommand \bibnamefont  [1]{#1}%
\providecommand \bibfnamefont [1]{#1}%
\providecommand \citenamefont [1]{#1}%
\providecommand \href@noop [0]{\@secondoftwo}%
\providecommand \href [0]{\begingroup \@sanitize@url \@href}%
\providecommand \@href[1]{\@@startlink{#1}\@@href}%
\providecommand \@@href[1]{\endgroup#1\@@endlink}%
\providecommand \@sanitize@url [0]{\catcode `\\12\catcode `\$12\catcode
  `\&12\catcode `\#12\catcode `\^12\catcode `\_12\catcode `\%12\relax}%
\providecommand \@@startlink[1]{}%
\providecommand \@@endlink[0]{}%
\providecommand \url  [0]{\begingroup\@sanitize@url \@url }%
\providecommand \@url [1]{\endgroup\@href {#1}{\urlprefix }}%
\providecommand \urlprefix  [0]{URL }%
\providecommand \Eprint [0]{\href }%
\providecommand \doibase [0]{http://dx.doi.org/}%
\providecommand \selectlanguage [0]{\@gobble}%
\providecommand \bibinfo  [0]{\@secondoftwo}%
\providecommand \bibfield  [0]{\@secondoftwo}%
\providecommand \translation [1]{[#1]}%
\providecommand \BibitemOpen [0]{}%
\providecommand \bibitemStop [0]{}%
\providecommand \bibitemNoStop [0]{.\EOS\space}%
\providecommand \EOS [0]{\spacefactor3000\relax}%
\providecommand \BibitemShut  [1]{\csname bibitem#1\endcsname}%
\let\auto@bib@innerbib\@empty
\bibitem [{\citenamefont {Cornia}\ and\ \citenamefont
  {Seneor}(2017)}]{01Cornia2017}%
  \BibitemOpen
  \bibfield  {author} {\bibinfo {author} {\bibfnamefont {A.}~\bibnamefont
  {Cornia}}\ and\ \bibinfo {author} {\bibfnamefont {P.}~\bibnamefont
  {Seneor}},\ }\href@noop {} {\bibfield  {journal} {\bibinfo  {journal} {Nature
  Materials}\ }\textbf {\bibinfo {volume} {16}},\ \bibinfo {pages} {505}
  (\bibinfo {year} {2017})}\BibitemShut {NoStop}%
\bibitem [{\citenamefont {Cinchetti}\ \emph {et~al.}(2017)\citenamefont
  {Cinchetti}, \citenamefont {Dediu},\ and\ \citenamefont
  {Hueso}}]{02Cinchetti2017}%
  \BibitemOpen
  \bibfield  {author} {\bibinfo {author} {\bibfnamefont {M.}~\bibnamefont
  {Cinchetti}}, \bibinfo {author} {\bibfnamefont {V.~A.}\ \bibnamefont
  {Dediu}}, \ and\ \bibinfo {author} {\bibfnamefont {L.~E.}\ \bibnamefont
  {Hueso}},\ }\href@noop {} {\bibfield  {journal} {\bibinfo  {journal} {Nature
  Materials}\ }\textbf {\bibinfo {volume} {16}},\ \bibinfo {pages} {507}
  (\bibinfo {year} {2017})}\BibitemShut {NoStop}%
\bibitem [{\citenamefont {Yu}\ \emph {et~al.}(2014)\citenamefont {Yu},
  \citenamefont {Kim}, \citenamefont {Lee},\ and\ \citenamefont
  {Oh}}]{03Yu2014}%
  \BibitemOpen
  \bibfield  {author} {\bibinfo {author} {\bibfnamefont {H.}~\bibnamefont
  {Yu}}, \bibinfo {author} {\bibfnamefont {D.~Y.}\ \bibnamefont {Kim}},
  \bibinfo {author} {\bibfnamefont {K.~J.}\ \bibnamefont {Lee}}, \ and\
  \bibinfo {author} {\bibfnamefont {J.~H.}\ \bibnamefont {Oh}},\ }\href
  {\doibase 10.1166/jnn.2014.9086} {\bibfield  {journal} {\bibinfo  {journal}
  {Journal of Nanoscience and Nanotechnology}\ }\textbf {\bibinfo {volume}
  {14}},\ \bibinfo {pages} {1282} (\bibinfo {year} {2014})}\BibitemShut
  {NoStop}%
\bibitem [{\citenamefont {Warner}\ \emph {et~al.}(2013)\citenamefont {Warner},
  \citenamefont {Din}, \citenamefont {Tupitsyn}, \citenamefont {Morley},
  \citenamefont {Stoneham}, \citenamefont {Gardener}, \citenamefont {Wu},
  \citenamefont {Fisher}, \citenamefont {Heutz}, \citenamefont {Kay},\ and\
  \citenamefont {Aeppli}}]{04Warner2013}%
  \BibitemOpen
  \bibfield  {author} {\bibinfo {author} {\bibfnamefont {M.}~\bibnamefont
  {Warner}}, \bibinfo {author} {\bibfnamefont {S.}~\bibnamefont {Din}},
  \bibinfo {author} {\bibfnamefont {I.~S.}\ \bibnamefont {Tupitsyn}}, \bibinfo
  {author} {\bibfnamefont {G.~W.}\ \bibnamefont {Morley}}, \bibinfo {author}
  {\bibfnamefont {A.~M.}\ \bibnamefont {Stoneham}}, \bibinfo {author}
  {\bibfnamefont {J.~A.}\ \bibnamefont {Gardener}}, \bibinfo {author}
  {\bibfnamefont {Z.}~\bibnamefont {Wu}}, \bibinfo {author} {\bibfnamefont
  {A.~J.}\ \bibnamefont {Fisher}}, \bibinfo {author} {\bibfnamefont
  {S.}~\bibnamefont {Heutz}}, \bibinfo {author} {\bibfnamefont {C.~W.~M.}\
  \bibnamefont {Kay}}, \ and\ \bibinfo {author} {\bibfnamefont
  {G.}~\bibnamefont {Aeppli}},\ }\href@noop {} {\bibfield  {journal} {\bibinfo
  {journal} {Nature}\ }\textbf {\bibinfo {volume} {503}},\ \bibinfo {pages}
  {504} (\bibinfo {year} {2013})}\BibitemShut {NoStop}%
\bibitem [{\citenamefont {Gambardella}\ \emph {et~al.}(2002)\citenamefont
  {Gambardella}, \citenamefont {Dallmeyer}, \citenamefont {Maiti},
  \citenamefont {Malagoli}, \citenamefont {Eberhardt}, \citenamefont {Kern},\
  and\ \citenamefont {Carbone}}]{05Gambardella2002}%
  \BibitemOpen
  \bibfield  {author} {\bibinfo {author} {\bibfnamefont {P.}~\bibnamefont
  {Gambardella}}, \bibinfo {author} {\bibfnamefont {A.}~\bibnamefont
  {Dallmeyer}}, \bibinfo {author} {\bibfnamefont {K.}~\bibnamefont {Maiti}},
  \bibinfo {author} {\bibfnamefont {M.~C.}\ \bibnamefont {Malagoli}}, \bibinfo
  {author} {\bibfnamefont {W.}~\bibnamefont {Eberhardt}}, \bibinfo {author}
  {\bibfnamefont {K.}~\bibnamefont {Kern}}, \ and\ \bibinfo {author}
  {\bibfnamefont {C.}~\bibnamefont {Carbone}},\ }\href {\doibase
  10.1038/416301a} {\bibfield  {journal} {\bibinfo  {journal} {Nature}\
  }\textbf {\bibinfo {volume} {416}},\ \bibinfo {pages} {301} (\bibinfo {year}
  {2002})}\BibitemShut {NoStop}%
\bibitem [{\citenamefont {Govenius}\ \emph {et~al.}(2016)\citenamefont
  {Govenius}, \citenamefont {Lake}, \citenamefont {Tan},\ and\ \citenamefont
  {M\"ott\"onen}}]{06PhysRevLett.117.030802}%
  \BibitemOpen
  \bibfield  {author} {\bibinfo {author} {\bibfnamefont {J.}~\bibnamefont
  {Govenius}}, \bibinfo {author} {\bibfnamefont {R.~E.}\ \bibnamefont {Lake}},
  \bibinfo {author} {\bibfnamefont {K.~Y.}\ \bibnamefont {Tan}}, \ and\
  \bibinfo {author} {\bibfnamefont {M.}~\bibnamefont {M\"ott\"onen}},\ }\href
  {\doibase 10.1103/PhysRevLett.117.030802} {\bibfield  {journal} {\bibinfo
  {journal} {Phys. Rev. Lett.}\ }\textbf {\bibinfo {volume} {117}},\ \bibinfo
  {pages} {030802} (\bibinfo {year} {2016})}\BibitemShut {NoStop}%
\bibitem [{\citenamefont {Gatteschi}\ \emph {et~al.}(2000)\citenamefont
  {Gatteschi}, \citenamefont {Sessoli},\ and\ \citenamefont
  {Cornia}}]{07Gatteschi2000}%
  \BibitemOpen
  \bibfield  {author} {\bibinfo {author} {\bibfnamefont {D.}~\bibnamefont
  {Gatteschi}}, \bibinfo {author} {\bibfnamefont {R.}~\bibnamefont {Sessoli}},
  \ and\ \bibinfo {author} {\bibfnamefont {A.}~\bibnamefont {Cornia}},\ }\href
  {\doibase 10.1039/A908254I} {\bibfield  {journal} {\bibinfo  {journal} {Chem.
  Commun.}\ ,\ \bibinfo {pages} {725}} (\bibinfo {year} {2000})}\BibitemShut
  {NoStop}%
\bibitem [{\citenamefont {Mannini}\ \emph {et~al.}(2009)\citenamefont
  {Mannini}, \citenamefont {Pineider}, \citenamefont {Sainctavit},
  \citenamefont {Danieli}, \citenamefont {Otero}, \citenamefont
  {Sciancalepore}, \citenamefont {Talarico}, \citenamefont {Arrio},
  \citenamefont {Cornia}, \citenamefont {Gatteschi},\ and\ \citenamefont
  {Sessoli}}]{08Mannini2009}%
  \BibitemOpen
  \bibfield  {author} {\bibinfo {author} {\bibfnamefont {M.}~\bibnamefont
  {Mannini}}, \bibinfo {author} {\bibfnamefont {F.}~\bibnamefont {Pineider}},
  \bibinfo {author} {\bibfnamefont {P.}~\bibnamefont {Sainctavit}}, \bibinfo
  {author} {\bibfnamefont {C.}~\bibnamefont {Danieli}}, \bibinfo {author}
  {\bibfnamefont {E.}~\bibnamefont {Otero}}, \bibinfo {author} {\bibfnamefont
  {C.}~\bibnamefont {Sciancalepore}}, \bibinfo {author} {\bibfnamefont {A.~M.}\
  \bibnamefont {Talarico}}, \bibinfo {author} {\bibfnamefont {M.-A.}\
  \bibnamefont {Arrio}}, \bibinfo {author} {\bibfnamefont {A.}~\bibnamefont
  {Cornia}}, \bibinfo {author} {\bibfnamefont {D.}~\bibnamefont {Gatteschi}}, \
  and\ \bibinfo {author} {\bibfnamefont {R.}~\bibnamefont {Sessoli}},\
  }\href@noop {} {\bibfield  {journal} {\bibinfo  {journal} {Nature Materials}\
  }\textbf {\bibinfo {volume} {8}},\ \bibinfo {pages} {194} (\bibinfo {year}
  {2009})}\BibitemShut {NoStop}%
\bibitem [{\citenamefont {Mannini}\ \emph {et~al.}(2005)\citenamefont
  {Mannini}, \citenamefont {Bonacchi}, \citenamefont {Zobbi}, \citenamefont
  {Piras}, \citenamefont {Speets}, \citenamefont {Caneschi}, \citenamefont
  {Cornia}, \citenamefont {Magnani}, \citenamefont {Ravoo}, \citenamefont
  {Reinhoudt}, \citenamefont {Sessoli},\ and\ \citenamefont
  {Gatteschi}}]{09Mannini2005}%
  \BibitemOpen
  \bibfield  {author} {\bibinfo {author} {\bibfnamefont {M.}~\bibnamefont
  {Mannini}}, \bibinfo {author} {\bibfnamefont {D.}~\bibnamefont {Bonacchi}},
  \bibinfo {author} {\bibfnamefont {L.}~\bibnamefont {Zobbi}}, \bibinfo
  {author} {\bibfnamefont {F.~M.}\ \bibnamefont {Piras}}, \bibinfo {author}
  {\bibfnamefont {E.~A.}\ \bibnamefont {Speets}}, \bibinfo {author}
  {\bibfnamefont {A.}~\bibnamefont {Caneschi}}, \bibinfo {author}
  {\bibfnamefont {A.}~\bibnamefont {Cornia}}, \bibinfo {author} {\bibfnamefont
  {A.}~\bibnamefont {Magnani}}, \bibinfo {author} {\bibfnamefont {B.~J.}\
  \bibnamefont {Ravoo}}, \bibinfo {author} {\bibfnamefont {D.~N.}\ \bibnamefont
  {Reinhoudt}}, \bibinfo {author} {\bibfnamefont {R.}~\bibnamefont {Sessoli}},
  \ and\ \bibinfo {author} {\bibfnamefont {D.}~\bibnamefont {Gatteschi}},\
  }\href {\doibase 10.1021/nl0508016} {\bibfield  {journal} {\bibinfo
  {journal} {Nano Letters}\ }\textbf {\bibinfo {volume} {5}},\ \bibinfo {pages}
  {1435} (\bibinfo {year} {2005})}\BibitemShut {NoStop}%
\bibitem [{\citenamefont {Sessoli}\ \emph {et~al.}(1993)\citenamefont
  {Sessoli}, \citenamefont {Gatteschi}, \citenamefont {Caneschi},\ and\
  \citenamefont {Novak}}]{10Sessoli1993}%
  \BibitemOpen
  \bibfield  {author} {\bibinfo {author} {\bibfnamefont {R.}~\bibnamefont
  {Sessoli}}, \bibinfo {author} {\bibfnamefont {D.}~\bibnamefont {Gatteschi}},
  \bibinfo {author} {\bibfnamefont {A.}~\bibnamefont {Caneschi}}, \ and\
  \bibinfo {author} {\bibfnamefont {M.~A.}\ \bibnamefont {Novak}},\ }\href
  {\doibase 10.1038/365141a0} {\bibfield  {journal} {\bibinfo  {journal}
  {Nature}\ }\textbf {\bibinfo {volume} {365}},\ \bibinfo {pages} {141}
  (\bibinfo {year} {1993})}\BibitemShut {NoStop}%
\bibitem [{\citenamefont {Woodruff}\ \emph {et~al.}(2013)\citenamefont
  {Woodruff}, \citenamefont {Winpenny},\ and\ \citenamefont
  {Layfield}}]{11Woodruff2013}%
  \BibitemOpen
  \bibfield  {author} {\bibinfo {author} {\bibfnamefont {D.~N.}\ \bibnamefont
  {Woodruff}}, \bibinfo {author} {\bibfnamefont {R.~E.~P.}\ \bibnamefont
  {Winpenny}}, \ and\ \bibinfo {author} {\bibfnamefont {R.~A.}\ \bibnamefont
  {Layfield}},\ }\href {\doibase 10.1021/cr400018q} {\bibfield  {journal}
  {\bibinfo  {journal} {Chemical Reviews}\ }\textbf {\bibinfo {volume} {113}},\
  \bibinfo {pages} {5110} (\bibinfo {year} {2013})}\BibitemShut {NoStop}%
\bibitem [{\citenamefont {Gatteschi}\ and\ \citenamefont
  {Sessoli}(2003)}]{12Gatteschi2003}%
  \BibitemOpen
  \bibfield  {author} {\bibinfo {author} {\bibfnamefont {D.}~\bibnamefont
  {Gatteschi}}\ and\ \bibinfo {author} {\bibfnamefont {R.}~\bibnamefont
  {Sessoli}},\ }\href {\doibase 10.1002/anie.200390099} {\bibfield  {journal}
  {\bibinfo  {journal} {Angewandte Chemie International Edition}\ }\textbf
  {\bibinfo {volume} {42}},\ \bibinfo {pages} {268} (\bibinfo {year}
  {2003})}\BibitemShut {NoStop}%
\bibitem [{\citenamefont {Thomas}\ \emph {et~al.}(1996)\citenamefont {Thomas},
  \citenamefont {Lionti}, \citenamefont {Ballou}, \citenamefont {Gatteschi},
  \citenamefont {Sessoli},\ and\ \citenamefont {Barbara}}]{13Thomas1996}%
  \BibitemOpen
  \bibfield  {author} {\bibinfo {author} {\bibfnamefont {L.}~\bibnamefont
  {Thomas}}, \bibinfo {author} {\bibfnamefont {F.}~\bibnamefont {Lionti}},
  \bibinfo {author} {\bibfnamefont {R.}~\bibnamefont {Ballou}}, \bibinfo
  {author} {\bibfnamefont {D.}~\bibnamefont {Gatteschi}}, \bibinfo {author}
  {\bibfnamefont {R.}~\bibnamefont {Sessoli}}, \ and\ \bibinfo {author}
  {\bibfnamefont {B.}~\bibnamefont {Barbara}},\ }\href {\doibase
  10.1038/383145a0} {\bibfield  {journal} {\bibinfo  {journal} {Nature}\
  }\textbf {\bibinfo {volume} {383}},\ \bibinfo {pages} {145} (\bibinfo {year}
  {1996})}\BibitemShut {NoStop}%
\bibitem [{\citenamefont {Friedman}\ \emph {et~al.}(1996)\citenamefont
  {Friedman}, \citenamefont {Sarachik}, \citenamefont {Tejada},\ and\
  \citenamefont {Ziolo}}]{14Friedman1996}%
  \BibitemOpen
  \bibfield  {author} {\bibinfo {author} {\bibfnamefont {J.~R.}\ \bibnamefont
  {Friedman}}, \bibinfo {author} {\bibfnamefont {M.~P.}\ \bibnamefont
  {Sarachik}}, \bibinfo {author} {\bibfnamefont {J.}~\bibnamefont {Tejada}}, \
  and\ \bibinfo {author} {\bibfnamefont {R.}~\bibnamefont {Ziolo}},\ }\href
  {\doibase 10.1103/PhysRevLett.76.3830} {\bibfield  {journal} {\bibinfo
  {journal} {Phys. Rev. Lett.}\ }\textbf {\bibinfo {volume} {76}},\ \bibinfo
  {pages} {3830} (\bibinfo {year} {1996})}\BibitemShut {NoStop}%
\bibitem [{\citenamefont {Ishikawa}\ \emph {et~al.}(2005)\citenamefont
  {Ishikawa}, \citenamefont {Sugita},\ and\ \citenamefont
  {Wernsdorfer}}]{15Ishikawa2005}%
  \BibitemOpen
  \bibfield  {author} {\bibinfo {author} {\bibfnamefont {N.}~\bibnamefont
  {Ishikawa}}, \bibinfo {author} {\bibfnamefont {M.}~\bibnamefont {Sugita}}, \
  and\ \bibinfo {author} {\bibfnamefont {W.}~\bibnamefont {Wernsdorfer}},\
  }\href {\doibase 10.1002/anie.200462638} {\bibfield  {journal} {\bibinfo
  {journal} {Angewandte Chemie International Edition}\ }\textbf {\bibinfo
  {volume} {44}},\ \bibinfo {pages} {2931} (\bibinfo {year}
  {2005})}\BibitemShut {NoStop}%
\bibitem [{\citenamefont {Ganzhorn}\ \emph {et~al.}(2016)\citenamefont
  {Ganzhorn}, \citenamefont {Klyatskaya}, \citenamefont {Ruben},\ and\
  \citenamefont {Wernsdorfer}}]{16Ganzhorn2016}%
  \BibitemOpen
  \bibfield  {author} {\bibinfo {author} {\bibfnamefont {M.}~\bibnamefont
  {Ganzhorn}}, \bibinfo {author} {\bibfnamefont {S.}~\bibnamefont
  {Klyatskaya}}, \bibinfo {author} {\bibfnamefont {M.}~\bibnamefont {Ruben}}, \
  and\ \bibinfo {author} {\bibfnamefont {W.}~\bibnamefont {Wernsdorfer}},\
  }\href {\doibase 10.1038/ncomms11443} {\bibfield  {journal} {\bibinfo
  {journal} {Nature Communications}\ }\textbf {\bibinfo {volume} {7}},\
  \bibinfo {pages} {11443} (\bibinfo {year} {2016})}\BibitemShut {NoStop}%
\bibitem [{\citenamefont {Urdampilleta}\ \emph {et~al.}(2011)\citenamefont
  {Urdampilleta}, \citenamefont {Klyatskaya}, \citenamefont {Cleuziou},
  \citenamefont {Ruben},\ and\ \citenamefont
  {Wernsdorfer}}]{17Urdampilleta2011}%
  \BibitemOpen
  \bibfield  {author} {\bibinfo {author} {\bibfnamefont {M.}~\bibnamefont
  {Urdampilleta}}, \bibinfo {author} {\bibfnamefont {S.}~\bibnamefont
  {Klyatskaya}}, \bibinfo {author} {\bibfnamefont {J.-P.}\ \bibnamefont
  {Cleuziou}}, \bibinfo {author} {\bibfnamefont {M.}~\bibnamefont {Ruben}}, \
  and\ \bibinfo {author} {\bibfnamefont {W.}~\bibnamefont {Wernsdorfer}},\
  }\href@noop {} {\bibfield  {journal} {\bibinfo  {journal} {Nature Materials}\
  }\textbf {\bibinfo {volume} {10}},\ \bibinfo {pages} {502} (\bibinfo {year}
  {2011})}\BibitemShut {NoStop}%
\bibitem [{\citenamefont {Ganzhorn}\ \emph {et~al.}(2013)\citenamefont
  {Ganzhorn}, \citenamefont {Klyatskaya}, \citenamefont {Ruben},\ and\
  \citenamefont {Wernsdorfer}}]{18Ganzhorn2013}%
  \BibitemOpen
  \bibfield  {author} {\bibinfo {author} {\bibfnamefont {M.}~\bibnamefont
  {Ganzhorn}}, \bibinfo {author} {\bibfnamefont {S.}~\bibnamefont
  {Klyatskaya}}, \bibinfo {author} {\bibfnamefont {M.}~\bibnamefont {Ruben}}, \
  and\ \bibinfo {author} {\bibfnamefont {W.}~\bibnamefont {Wernsdorfer}},\
  }\href@noop {} {\bibfield  {journal} {\bibinfo  {journal} {Nature
  Nanotechnology}\ }\textbf {\bibinfo {volume} {8}},\ \bibinfo {pages} {165}
  (\bibinfo {year} {2013})}\BibitemShut {NoStop}%
\bibitem [{\citenamefont {W{\"{a}}ckerlin}\ \emph {et~al.}(2016)\citenamefont
  {W{\"{a}}ckerlin}, \citenamefont {Donati}, \citenamefont {Singha},
  \citenamefont {Baltic}, \citenamefont {Rusponi}, \citenamefont {Diller},
  \citenamefont {Patthey}, \citenamefont {Pivetta}, \citenamefont {Lan},
  \citenamefont {Klyatskaya}, \citenamefont {Ruben}, \citenamefont {Brune},\
  and\ \citenamefont {Dreiser}}]{19Wackerlin2016}%
  \BibitemOpen
  \bibfield  {author} {\bibinfo {author} {\bibfnamefont {C.}~\bibnamefont
  {W{\"{a}}ckerlin}}, \bibinfo {author} {\bibfnamefont {F.}~\bibnamefont
  {Donati}}, \bibinfo {author} {\bibfnamefont {A.}~\bibnamefont {Singha}},
  \bibinfo {author} {\bibfnamefont {R.}~\bibnamefont {Baltic}}, \bibinfo
  {author} {\bibfnamefont {S.}~\bibnamefont {Rusponi}}, \bibinfo {author}
  {\bibfnamefont {K.}~\bibnamefont {Diller}}, \bibinfo {author} {\bibfnamefont
  {F.}~\bibnamefont {Patthey}}, \bibinfo {author} {\bibfnamefont
  {M.}~\bibnamefont {Pivetta}}, \bibinfo {author} {\bibfnamefont
  {Y.}~\bibnamefont {Lan}}, \bibinfo {author} {\bibfnamefont {S.}~\bibnamefont
  {Klyatskaya}}, \bibinfo {author} {\bibfnamefont {M.}~\bibnamefont {Ruben}},
  \bibinfo {author} {\bibfnamefont {H.}~\bibnamefont {Brune}}, \ and\ \bibinfo
  {author} {\bibfnamefont {J.}~\bibnamefont {Dreiser}},\ }\href {\doibase
  10.1002/adma.201506305} {\bibfield  {journal} {\bibinfo  {journal} {Advanced
  Materials}\ }\textbf {\bibinfo {volume} {28}},\ \bibinfo {pages} {5195}
  (\bibinfo {year} {2016})}\BibitemShut {NoStop}%
\bibitem [{\citenamefont {Ara}\ \emph {et~al.}(2019)\citenamefont {Ara},
  \citenamefont {Oka}, \citenamefont {Sainoo}, \citenamefont {Katoh},
  \citenamefont {Yamashita},\ and\ \citenamefont {Komeda}}]{20Ara2019}%
  \BibitemOpen
  \bibfield  {author} {\bibinfo {author} {\bibfnamefont {F.}~\bibnamefont
  {Ara}}, \bibinfo {author} {\bibfnamefont {H.}~\bibnamefont {Oka}}, \bibinfo
  {author} {\bibfnamefont {Y.}~\bibnamefont {Sainoo}}, \bibinfo {author}
  {\bibfnamefont {K.}~\bibnamefont {Katoh}}, \bibinfo {author} {\bibfnamefont
  {M.}~\bibnamefont {Yamashita}}, \ and\ \bibinfo {author} {\bibfnamefont
  {T.}~\bibnamefont {Komeda}},\ }\href {\doibase 10.1063/1.5079964} {\bibfield
  {journal} {\bibinfo  {journal} {Journal of Applied Physics}\ }\textbf
  {\bibinfo {volume} {125}},\ \bibinfo {pages} {183901} (\bibinfo {year}
  {2019})}\BibitemShut {NoStop}%
\bibitem [{\citenamefont {Papageorgiou}\ \emph {et~al.}(2004)\citenamefont
  {Papageorgiou}, \citenamefont {Salomon}, \citenamefont {Angot}, \citenamefont
  {Layet}, \citenamefont {Giovanelli},\ and\ \citenamefont
  {Le~Lay}}]{21Papageorgiou2004}%
  \BibitemOpen
  \bibfield  {author} {\bibinfo {author} {\bibfnamefont {N.}~\bibnamefont
  {Papageorgiou}}, \bibinfo {author} {\bibfnamefont {E.}~\bibnamefont
  {Salomon}}, \bibinfo {author} {\bibfnamefont {T.}~\bibnamefont {Angot}},
  \bibinfo {author} {\bibfnamefont {J.-M.}\ \bibnamefont {Layet}}, \bibinfo
  {author} {\bibfnamefont {L.}~\bibnamefont {Giovanelli}}, \ and\ \bibinfo
  {author} {\bibfnamefont {G.}~\bibnamefont {Le~Lay}},\ }\href {\doibase
  https://doi.org/10.1016/j.progsurf.2005.01.001} {\bibfield  {journal}
  {\bibinfo  {journal} {Progress in Surface Science}\ }\textbf {\bibinfo
  {volume} {77}},\ \bibinfo {pages} {139} (\bibinfo {year} {2004})}\BibitemShut
  {NoStop}%
\bibitem [{\citenamefont {Angione}\ \emph {et~al.}(2011)\citenamefont
  {Angione}, \citenamefont {Pilolli}, \citenamefont {Cotrone}, \citenamefont
  {Magliulo}, \citenamefont {Mallardi}, \citenamefont {Palazzo}, \citenamefont
  {Sabbatini}, \citenamefont {Fine}, \citenamefont {Dodabalapur}, \citenamefont
  {Cioffi},\ and\ \citenamefont {Torsi}}]{22Angione2011}%
  \BibitemOpen
  \bibfield  {author} {\bibinfo {author} {\bibfnamefont {M.~D.}\ \bibnamefont
  {Angione}}, \bibinfo {author} {\bibfnamefont {R.}~\bibnamefont {Pilolli}},
  \bibinfo {author} {\bibfnamefont {S.}~\bibnamefont {Cotrone}}, \bibinfo
  {author} {\bibfnamefont {M.}~\bibnamefont {Magliulo}}, \bibinfo {author}
  {\bibfnamefont {A.}~\bibnamefont {Mallardi}}, \bibinfo {author}
  {\bibfnamefont {G.}~\bibnamefont {Palazzo}}, \bibinfo {author} {\bibfnamefont
  {L.}~\bibnamefont {Sabbatini}}, \bibinfo {author} {\bibfnamefont
  {D.}~\bibnamefont {Fine}}, \bibinfo {author} {\bibfnamefont {A.}~\bibnamefont
  {Dodabalapur}}, \bibinfo {author} {\bibfnamefont {N.}~\bibnamefont {Cioffi}},
  \ and\ \bibinfo {author} {\bibfnamefont {L.}~\bibnamefont {Torsi}},\ }\href
  {\doibase https://doi.org/10.1016/S1369-7021(11)70187-0} {\bibfield
  {journal} {\bibinfo  {journal} {Materials Today}\ }\textbf {\bibinfo {volume}
  {14}},\ \bibinfo {pages} {424} (\bibinfo {year} {2011})}\BibitemShut
  {NoStop}%
\bibitem [{\citenamefont {Bao}\ \emph {et~al.}(1996)\citenamefont {Bao},
  \citenamefont {Lovinger},\ and\ \citenamefont {Dodabalapur}}]{23Bao1996}%
  \BibitemOpen
  \bibfield  {author} {\bibinfo {author} {\bibfnamefont {Z.}~\bibnamefont
  {Bao}}, \bibinfo {author} {\bibfnamefont {A.~J.}\ \bibnamefont {Lovinger}}, \
  and\ \bibinfo {author} {\bibfnamefont {A.}~\bibnamefont {Dodabalapur}},\
  }\href {\doibase 10.1063/1.116841} {\bibfield  {journal} {\bibinfo  {journal}
  {Applied Physics Letters}\ }\textbf {\bibinfo {volume} {69}},\ \bibinfo
  {pages} {3066} (\bibinfo {year} {1996})}\BibitemShut {NoStop}%
\bibitem [{\citenamefont {Nguyen}(2011)}]{24Nguyen2011}%
  \BibitemOpen
  \bibfield  {author} {\bibinfo {author} {\bibfnamefont {T.-P.}\ \bibnamefont
  {Nguyen}},\ }\href {\doibase https://doi.org/10.1016/j.surfcoat.2011.07.010}
  {\bibfield  {journal} {\bibinfo  {journal} {Surface and Coatings Technology}\
  }\textbf {\bibinfo {volume} {206}},\ \bibinfo {pages} {742} (\bibinfo {year}
  {2011})}\BibitemShut {NoStop}%
\bibitem [{\citenamefont {Wu}\ \emph {et~al.}(2004)\citenamefont {Wu},
  \citenamefont {Nazin}, \citenamefont {Chen}, \citenamefont {Qiu},\ and\
  \citenamefont {Ho}}]{25Wu2004}%
  \BibitemOpen
  \bibfield  {author} {\bibinfo {author} {\bibfnamefont {S.~W.}\ \bibnamefont
  {Wu}}, \bibinfo {author} {\bibfnamefont {G.~V.}\ \bibnamefont {Nazin}},
  \bibinfo {author} {\bibfnamefont {X.}~\bibnamefont {Chen}}, \bibinfo {author}
  {\bibfnamefont {X.~H.}\ \bibnamefont {Qiu}}, \ and\ \bibinfo {author}
  {\bibfnamefont {W.}~\bibnamefont {Ho}},\ }\href {\doibase
  10.1103/PhysRevLett.93.236802} {\bibfield  {journal} {\bibinfo  {journal}
  {Phys. Rev. Lett.}\ }\textbf {\bibinfo {volume} {93}},\ \bibinfo {pages}
  {236802} (\bibinfo {year} {2004})}\BibitemShut {NoStop}%
\bibitem [{\citenamefont {Peumans}\ and\ \citenamefont
  {Forrest}(2001)}]{26Peumans2001}%
  \BibitemOpen
  \bibfield  {author} {\bibinfo {author} {\bibfnamefont {P.}~\bibnamefont
  {Peumans}}\ and\ \bibinfo {author} {\bibfnamefont {S.~R.}\ \bibnamefont
  {Forrest}},\ }\href {\doibase 10.1063/1.1384001} {\bibfield  {journal}
  {\bibinfo  {journal} {Applied Physics Letters}\ }\textbf {\bibinfo {volume}
  {79}},\ \bibinfo {pages} {126} (\bibinfo {year} {2001})}\BibitemShut
  {NoStop}%
\bibitem [{\citenamefont {Lu}\ \emph {et~al.}(1996)\citenamefont {Lu},
  \citenamefont {Hipps}, \citenamefont {Wang},\ and\ \citenamefont
  {Mazur}}]{27Lu1996}%
  \BibitemOpen
  \bibfield  {author} {\bibinfo {author} {\bibfnamefont {X.}~\bibnamefont
  {Lu}}, \bibinfo {author} {\bibfnamefont {K.~W.}\ \bibnamefont {Hipps}},
  \bibinfo {author} {\bibfnamefont {X.~D.}\ \bibnamefont {Wang}}, \ and\
  \bibinfo {author} {\bibfnamefont {U.}~\bibnamefont {Mazur}},\ }\href
  {\doibase 10.1021/ja960874e} {\bibfield  {journal} {\bibinfo  {journal}
  {Journal of the American Chemical Society}\ }\textbf {\bibinfo {volume}
  {118}},\ \bibinfo {pages} {7197} (\bibinfo {year} {1996})}\BibitemShut
  {NoStop}%
\bibitem [{\citenamefont {Nazin}\ \emph {et~al.}(2003)\citenamefont {Nazin},
  \citenamefont {Qiu},\ and\ \citenamefont {Ho}}]{28Nazin2003}%
  \BibitemOpen
  \bibfield  {author} {\bibinfo {author} {\bibfnamefont {G.~V.}\ \bibnamefont
  {Nazin}}, \bibinfo {author} {\bibfnamefont {X.~H.}\ \bibnamefont {Qiu}}, \
  and\ \bibinfo {author} {\bibfnamefont {W.}~\bibnamefont {Ho}},\ }\href
  {\doibase 10.1126/science.1088971} {\bibfield  {journal} {\bibinfo  {journal}
  {Science}\ }\textbf {\bibinfo {volume} {302}},\ \bibinfo {pages} {77}
  (\bibinfo {year} {2003})}\BibitemShut {NoStop}%
\bibitem [{\citenamefont {Zhao}\ \emph {et~al.}(2005)\citenamefont {Zhao},
  \citenamefont {Li}, \citenamefont {Chen}, \citenamefont {Xiang},
  \citenamefont {Wang}, \citenamefont {Pan}, \citenamefont {Wang},
  \citenamefont {Xiao}, \citenamefont {Yang}, \citenamefont {Hou},\ and\
  \citenamefont {Zhu}}]{29Zhao2005}%
  \BibitemOpen
  \bibfield  {author} {\bibinfo {author} {\bibfnamefont {A.}~\bibnamefont
  {Zhao}}, \bibinfo {author} {\bibfnamefont {Q.}~\bibnamefont {Li}}, \bibinfo
  {author} {\bibfnamefont {L.}~\bibnamefont {Chen}}, \bibinfo {author}
  {\bibfnamefont {H.}~\bibnamefont {Xiang}}, \bibinfo {author} {\bibfnamefont
  {W.}~\bibnamefont {Wang}}, \bibinfo {author} {\bibfnamefont {S.}~\bibnamefont
  {Pan}}, \bibinfo {author} {\bibfnamefont {B.}~\bibnamefont {Wang}}, \bibinfo
  {author} {\bibfnamefont {X.}~\bibnamefont {Xiao}}, \bibinfo {author}
  {\bibfnamefont {J.}~\bibnamefont {Yang}}, \bibinfo {author} {\bibfnamefont
  {J.~G.}\ \bibnamefont {Hou}}, \ and\ \bibinfo {author} {\bibfnamefont
  {Q.}~\bibnamefont {Zhu}},\ }\href {\doibase 10.1126/science.1113449}
  {\bibfield  {journal} {\bibinfo  {journal} {Science}\ }\textbf {\bibinfo
  {volume} {309}},\ \bibinfo {pages} {1542} (\bibinfo {year}
  {2005})}\BibitemShut {NoStop}%
\bibitem [{\citenamefont {Zhao}\ \emph {et~al.}(2008)\citenamefont {Zhao},
  \citenamefont {Hu}, \citenamefont {Wang}, \citenamefont {Xiao}, \citenamefont
  {Yang},\ and\ \citenamefont {Hou}}]{30Zhao2008}%
  \BibitemOpen
  \bibfield  {author} {\bibinfo {author} {\bibfnamefont {A.}~\bibnamefont
  {Zhao}}, \bibinfo {author} {\bibfnamefont {Z.}~\bibnamefont {Hu}}, \bibinfo
  {author} {\bibfnamefont {B.}~\bibnamefont {Wang}}, \bibinfo {author}
  {\bibfnamefont {X.}~\bibnamefont {Xiao}}, \bibinfo {author} {\bibfnamefont
  {J.}~\bibnamefont {Yang}}, \ and\ \bibinfo {author} {\bibfnamefont {J.~G.}\
  \bibnamefont {Hou}},\ }\href {\doibase 10.1063/1.2940338} {\bibfield
  {journal} {\bibinfo  {journal} {The Journal of Chemical Physics}\ }\textbf
  {\bibinfo {volume} {128}},\ \bibinfo {pages} {234705} (\bibinfo {year}
  {2008})}\BibitemShut {NoStop}%
\bibitem [{\citenamefont {Chen}\ \emph {et~al.}(2007)\citenamefont {Chen},
  \citenamefont {Hu}, \citenamefont {Zhao}, \citenamefont {Wang}, \citenamefont
  {Luo}, \citenamefont {Yang},\ and\ \citenamefont {Hou}}]{31Chen2007}%
  \BibitemOpen
  \bibfield  {author} {\bibinfo {author} {\bibfnamefont {L.}~\bibnamefont
  {Chen}}, \bibinfo {author} {\bibfnamefont {Z.}~\bibnamefont {Hu}}, \bibinfo
  {author} {\bibfnamefont {A.}~\bibnamefont {Zhao}}, \bibinfo {author}
  {\bibfnamefont {B.}~\bibnamefont {Wang}}, \bibinfo {author} {\bibfnamefont
  {Y.}~\bibnamefont {Luo}}, \bibinfo {author} {\bibfnamefont {J.}~\bibnamefont
  {Yang}}, \ and\ \bibinfo {author} {\bibfnamefont {J.~G.}\ \bibnamefont
  {Hou}},\ }\href {\doibase 10.1103/PhysRevLett.99.146803} {\bibfield
  {journal} {\bibinfo  {journal} {Phys. Rev. Lett.}\ }\textbf {\bibinfo
  {volume} {99}},\ \bibinfo {pages} {146803} (\bibinfo {year}
  {2007})}\BibitemShut {NoStop}%
\bibitem [{\citenamefont {Villemin}\ \emph {et~al.}(2001)\citenamefont
  {Villemin}, \citenamefont {Hammadi}, \citenamefont {Hachemi},\ and\
  \citenamefont {Bar}}]{32Villemin2001}%
  \BibitemOpen
  \bibfield  {author} {\bibinfo {author} {\bibfnamefont {D.}~\bibnamefont
  {Villemin}}, \bibinfo {author} {\bibfnamefont {M.}~\bibnamefont {Hammadi}},
  \bibinfo {author} {\bibfnamefont {M.}~\bibnamefont {Hachemi}}, \ and\
  \bibinfo {author} {\bibfnamefont {N.}~\bibnamefont {Bar}},\ }\href {\doibase
  10.3390/61000831} {\bibfield  {journal} {\bibinfo  {journal} {Molecules}\
  }\textbf {\bibinfo {volume} {6}},\ \bibinfo {pages} {831} (\bibinfo {year}
  {2001})}\BibitemShut {NoStop}%
\bibitem [{\citenamefont {Gargiani}\ \emph
  {et~al.}(2010{\natexlab{a}})\citenamefont {Gargiani}, \citenamefont
  {Angelucci}, \citenamefont {Mariani},\ and\ \citenamefont
  {Betti}}]{33Gargiani2010}%
  \BibitemOpen
  \bibfield  {author} {\bibinfo {author} {\bibfnamefont {P.}~\bibnamefont
  {Gargiani}}, \bibinfo {author} {\bibfnamefont {M.}~\bibnamefont {Angelucci}},
  \bibinfo {author} {\bibfnamefont {C.}~\bibnamefont {Mariani}}, \ and\
  \bibinfo {author} {\bibfnamefont {M.~G.}\ \bibnamefont {Betti}},\ }\href
  {\doibase 10.1103/PhysRevB.81.085412} {\bibfield  {journal} {\bibinfo
  {journal} {Phys. Rev. B}\ }\textbf {\bibinfo {volume} {81}},\ \bibinfo
  {pages} {85412} (\bibinfo {year} {2010}{\natexlab{a}})}\BibitemShut {NoStop}%
\bibitem [{\citenamefont {K{\"{o}}nig}\ \emph {et~al.}(2009)\citenamefont
  {K{\"{o}}nig}, \citenamefont {Roth}, \citenamefont {Kraus},\ and\
  \citenamefont {Knupfer}}]{34Konig2009}%
  \BibitemOpen
  \bibfield  {author} {\bibinfo {author} {\bibfnamefont {A.}~\bibnamefont
  {K{\"{o}}nig}}, \bibinfo {author} {\bibfnamefont {F.}~\bibnamefont {Roth}},
  \bibinfo {author} {\bibfnamefont {R.}~\bibnamefont {Kraus}}, \ and\ \bibinfo
  {author} {\bibfnamefont {M.}~\bibnamefont {Knupfer}},\ }\href {\doibase
  10.1063/1.3146812} {\bibfield  {journal} {\bibinfo  {journal} {The Journal of
  Chemical Physics}\ }\textbf {\bibinfo {volume} {130}},\ \bibinfo {pages}
  {214503} (\bibinfo {year} {2009})}\BibitemShut {NoStop}%
\bibitem [{\citenamefont {Gargiani}\ \emph
  {et~al.}(2010{\natexlab{b}})\citenamefont {Gargiani}, \citenamefont
  {Calabrese}, \citenamefont {Mariani},\ and\ \citenamefont
  {Betti}}]{35Gargiani2010a}%
  \BibitemOpen
  \bibfield  {author} {\bibinfo {author} {\bibfnamefont {P.}~\bibnamefont
  {Gargiani}}, \bibinfo {author} {\bibfnamefont {A.}~\bibnamefont {Calabrese}},
  \bibinfo {author} {\bibfnamefont {C.}~\bibnamefont {Mariani}}, \ and\
  \bibinfo {author} {\bibfnamefont {M.~G.}\ \bibnamefont {Betti}},\ }\href
  {\doibase 10.1021/jp103946v} {\bibfield  {journal} {\bibinfo  {journal} {The
  Journal of Physical Chemistry C}\ }\textbf {\bibinfo {volume} {114}},\
  \bibinfo {pages} {12258} (\bibinfo {year} {2010}{\natexlab{b}})}\BibitemShut
  {NoStop}%
\bibitem [{\citenamefont {Hu}\ \emph {et~al.}(2008)\citenamefont {Hu},
  \citenamefont {Chen}, \citenamefont {Zhao}, \citenamefont {Li}, \citenamefont
  {Wang}, \citenamefont {Yang},\ and\ \citenamefont {Hou}}]{36Hu2008}%
  \BibitemOpen
  \bibfield  {author} {\bibinfo {author} {\bibfnamefont {Z.}~\bibnamefont
  {Hu}}, \bibinfo {author} {\bibfnamefont {L.}~\bibnamefont {Chen}}, \bibinfo
  {author} {\bibfnamefont {A.}~\bibnamefont {Zhao}}, \bibinfo {author}
  {\bibfnamefont {Z.}~\bibnamefont {Li}}, \bibinfo {author} {\bibfnamefont
  {B.}~\bibnamefont {Wang}}, \bibinfo {author} {\bibfnamefont {J.}~\bibnamefont
  {Yang}}, \ and\ \bibinfo {author} {\bibfnamefont {J.~G.}\ \bibnamefont
  {Hou}},\ }\href {\doibase 10.1021/jp8065508} {\bibfield  {journal} {\bibinfo
  {journal} {The Journal of Physical Chemistry C}\ }\textbf {\bibinfo {volume}
  {112}},\ \bibinfo {pages} {15603} (\bibinfo {year} {2008})}\BibitemShut
  {NoStop}%
\bibitem [{\citenamefont {Nardi}\ \emph {et~al.}(2013)\citenamefont {Nardi},
  \citenamefont {Detto}, \citenamefont {Aversa}, \citenamefont {Verucchi},
  \citenamefont {Salviati}, \citenamefont {Iannotta},\ and\ \citenamefont
  {Casarin}}]{37Nardi2013}%
  \BibitemOpen
  \bibfield  {author} {\bibinfo {author} {\bibfnamefont {M.~V.}\ \bibnamefont
  {Nardi}}, \bibinfo {author} {\bibfnamefont {F.}~\bibnamefont {Detto}},
  \bibinfo {author} {\bibfnamefont {L.}~\bibnamefont {Aversa}}, \bibinfo
  {author} {\bibfnamefont {R.}~\bibnamefont {Verucchi}}, \bibinfo {author}
  {\bibfnamefont {G.}~\bibnamefont {Salviati}}, \bibinfo {author}
  {\bibfnamefont {S.}~\bibnamefont {Iannotta}}, \ and\ \bibinfo {author}
  {\bibfnamefont {M.}~\bibnamefont {Casarin}},\ }\href {\doibase
  10.1039/C3CP51224J} {\bibfield  {journal} {\bibinfo  {journal} {Phys. Chem.
  Chem. Phys.}\ }\textbf {\bibinfo {volume} {15}},\ \bibinfo {pages} {12864}
  (\bibinfo {year} {2013})}\BibitemShut {NoStop}%
\bibitem [{\citenamefont {Kresse}\ and\ \citenamefont
  {Furthm\"uller}(1996)}]{38Kresse}%
  \BibitemOpen
  \bibfield  {author} {\bibinfo {author} {\bibfnamefont {G.}~\bibnamefont
  {Kresse}}\ and\ \bibinfo {author} {\bibfnamefont {J.}~\bibnamefont
  {Furthm\"uller}},\ }\href {\doibase 10.1103/PhysRevB.54.11169} {\bibfield
  {journal} {\bibinfo  {journal} {Phys. Rev. B}\ }\textbf {\bibinfo {volume}
  {54}},\ \bibinfo {pages} {11169} (\bibinfo {year} {1996})}\BibitemShut
  {NoStop}%
\bibitem [{\citenamefont {Perdew}\ \emph
  {et~al.}(1996{\natexlab{a}})\citenamefont {Perdew}, \citenamefont {Burke},\
  and\ \citenamefont {Ernzerhof}}]{39PerdewBurke}%
  \BibitemOpen
  \bibfield  {author} {\bibinfo {author} {\bibfnamefont {J.~P.}\ \bibnamefont
  {Perdew}}, \bibinfo {author} {\bibfnamefont {K.}~\bibnamefont {Burke}}, \
  and\ \bibinfo {author} {\bibfnamefont {M.}~\bibnamefont {Ernzerhof}},\ }\href
  {\doibase 10.1103/PhysRevLett.77.3865} {\bibfield  {journal} {\bibinfo
  {journal} {Phys. Rev. Lett.}\ }\textbf {\bibinfo {volume} {77}},\ \bibinfo
  {pages} {3865} (\bibinfo {year} {1996}{\natexlab{a}})}\BibitemShut {NoStop}%
\bibitem [{\citenamefont {Bl\"ochl}(1994)}]{40PAW}%
  \BibitemOpen
  \bibfield  {author} {\bibinfo {author} {\bibfnamefont {P.~E.}\ \bibnamefont
  {Bl\"ochl}},\ }\href {\doibase 10.1103/PhysRevB.50.17953} {\bibfield
  {journal} {\bibinfo  {journal} {Phys. Rev. B}\ }\textbf {\bibinfo {volume}
  {50}},\ \bibinfo {pages} {17953} (\bibinfo {year} {1994})}\BibitemShut
  {NoStop}%
\bibitem [{\citenamefont {Perdew}\ \emph
  {et~al.}(1996{\natexlab{b}})\citenamefont {Perdew}, \citenamefont
  {Ernzerhof},\ and\ \citenamefont {Burke}}]{41Perdew1996}%
  \BibitemOpen
  \bibfield  {author} {\bibinfo {author} {\bibfnamefont {J.~P.}\ \bibnamefont
  {Perdew}}, \bibinfo {author} {\bibfnamefont {M.}~\bibnamefont {Ernzerhof}}, \
  and\ \bibinfo {author} {\bibfnamefont {K.}~\bibnamefont {Burke}},\ }\href
  {\doibase 10.1063/1.472933} {\bibfield  {journal} {\bibinfo  {journal} {The
  Journal of Chemical Physics}\ }\textbf {\bibinfo {volume} {105}},\ \bibinfo
  {pages} {9982} (\bibinfo {year} {1996}{\natexlab{b}})}\BibitemShut {NoStop}%
\bibitem [{\citenamefont {Wang}\ \emph {et~al.}(2019)\citenamefont {Wang},
  \citenamefont {Li},\ and\ \citenamefont {Yang}}]{42Wang2019}%
  \BibitemOpen
  \bibfield  {author} {\bibinfo {author} {\bibfnamefont {Y.}~\bibnamefont
  {Wang}}, \bibinfo {author} {\bibfnamefont {X.}~\bibnamefont {Li}}, \ and\
  \bibinfo {author} {\bibfnamefont {J.}~\bibnamefont {Yang}},\ }\href {\doibase
  10.1039/c8cp07091a} {\bibfield  {journal} {\bibinfo  {journal} {Physical
  Chemistry Chemical Physics}\ }\textbf {\bibinfo {volume} {21}},\ \bibinfo
  {pages} {5424} (\bibinfo {year} {2019})}\BibitemShut {NoStop}%
\bibitem [{\citenamefont {Choi}\ \emph {et~al.}(2016)\citenamefont {Choi},
  \citenamefont {Robles}, \citenamefont {Gauyacq}, \citenamefont {Ternes},
  \citenamefont {Loth},\ and\ \citenamefont {Lorente}}]{43Choi2016}%
  \BibitemOpen
  \bibfield  {author} {\bibinfo {author} {\bibfnamefont {D.-J.}\ \bibnamefont
  {Choi}}, \bibinfo {author} {\bibfnamefont {R.}~\bibnamefont {Robles}},
  \bibinfo {author} {\bibfnamefont {J.-P.}\ \bibnamefont {Gauyacq}}, \bibinfo
  {author} {\bibfnamefont {M.}~\bibnamefont {Ternes}}, \bibinfo {author}
  {\bibfnamefont {S.}~\bibnamefont {Loth}}, \ and\ \bibinfo {author}
  {\bibfnamefont {N.}~\bibnamefont {Lorente}},\ }\href {\doibase
  10.1103/PhysRevB.94.085406} {\bibfield  {journal} {\bibinfo  {journal} {Phys.
  Rev. B}\ }\textbf {\bibinfo {volume} {94}},\ \bibinfo {pages} {085406}
  (\bibinfo {year} {2016})}\BibitemShut {NoStop}%
\bibitem [{\citenamefont {Czelej}\ \emph
  {et~al.}(2018{\natexlab{a}})\citenamefont {Czelej}, \citenamefont
  {{\'{C}}wieka}, \citenamefont {{\'{S}}piewak},\ and\ \citenamefont
  {Kurzyd{\l}owski}}]{57CzelejTi}%
  \BibitemOpen
  \bibfield  {author} {\bibinfo {author} {\bibfnamefont {K.}~\bibnamefont
  {Czelej}}, \bibinfo {author} {\bibfnamefont {K.}~\bibnamefont
  {{\'{C}}wieka}}, \bibinfo {author} {\bibfnamefont {P.}~\bibnamefont
  {{\'{S}}piewak}}, \ and\ \bibinfo {author} {\bibfnamefont {K.~J.}\
  \bibnamefont {Kurzyd{\l}owski}},\ }\href {\doibase 10.1039/C8TC00097B}
  {\bibfield  {journal} {\bibinfo  {journal} {J. Mater. Chem. C}\ }\textbf
  {\bibinfo {volume} {6}},\ \bibinfo {pages} {5261} (\bibinfo {year}
  {2018}{\natexlab{a}})}\BibitemShut {NoStop}%
\bibitem [{\citenamefont {Czelej}\ \emph
  {et~al.}(2018{\natexlab{b}})\citenamefont {Czelej}, \citenamefont {Zem{\l}a},
  \citenamefont {Kami{\'{n}}ska}, \citenamefont {{\'{S}}piewak},\ and\
  \citenamefont {Kurzyd{\l}owski}}]{58Czelej2018a}%
  \BibitemOpen
  \bibfield  {author} {\bibinfo {author} {\bibfnamefont {K.}~\bibnamefont
  {Czelej}}, \bibinfo {author} {\bibfnamefont {M.~R.}\ \bibnamefont
  {Zem{\l}a}}, \bibinfo {author} {\bibfnamefont {P.}~\bibnamefont
  {Kami{\'{n}}ska}}, \bibinfo {author} {\bibfnamefont {P.}~\bibnamefont
  {{\'{S}}piewak}}, \ and\ \bibinfo {author} {\bibfnamefont {K.~J.}\
  \bibnamefont {Kurzyd{\l}owski}},\ }\href {\doibase
  10.1103/PhysRevB.98.075208} {\bibfield  {journal} {\bibinfo  {journal}
  {Physical Review B}\ }\textbf {\bibinfo {volume} {98}},\ \bibinfo {pages}
  {075208} (\bibinfo {year} {2018}{\natexlab{b}})}\BibitemShut {NoStop}%
\bibitem [{\citenamefont {Czelej}\ \emph
  {et~al.}(2018{\natexlab{c}})\citenamefont {Czelej}, \citenamefont {Zem{\l}a},
  \citenamefont {{\'{S}}piewak},\ and\ \citenamefont
  {Kurzyd{\l}owski}}]{59Czelej2018}%
  \BibitemOpen
  \bibfield  {author} {\bibinfo {author} {\bibfnamefont {K.}~\bibnamefont
  {Czelej}}, \bibinfo {author} {\bibfnamefont {M.~R.}\ \bibnamefont
  {Zem{\l}a}}, \bibinfo {author} {\bibfnamefont {P.}~\bibnamefont
  {{\'{S}}piewak}}, \ and\ \bibinfo {author} {\bibfnamefont {K.~J.}\
  \bibnamefont {Kurzyd{\l}owski}},\ }\href {\doibase
  10.1103/PhysRevB.98.235111} {\bibfield  {journal} {\bibinfo  {journal}
  {Physical Review B}\ }\textbf {\bibinfo {volume} {98}},\ \bibinfo {pages}
  {235111} (\bibinfo {year} {2018}{\natexlab{c}})}\BibitemShut {NoStop}%
\bibitem [{\citenamefont {Bidermane}\ \emph {et~al.}(2015)\citenamefont
  {Bidermane}, \citenamefont {L{\"{u}}der}, \citenamefont {Totani},
  \citenamefont {Grazioli}, \citenamefont {de~Simone}, \citenamefont {Coreno},
  \citenamefont {Kivim{\"{a}}ki}, \citenamefont {{\AA}hlund}, \citenamefont
  {Lozzi}, \citenamefont {Brena},\ and\ \citenamefont
  {Puglia}}]{44Bidermane2015}%
  \BibitemOpen
  \bibfield  {author} {\bibinfo {author} {\bibfnamefont {I.}~\bibnamefont
  {Bidermane}}, \bibinfo {author} {\bibfnamefont {J.}~\bibnamefont
  {L{\"{u}}der}}, \bibinfo {author} {\bibfnamefont {R.}~\bibnamefont {Totani}},
  \bibinfo {author} {\bibfnamefont {C.}~\bibnamefont {Grazioli}}, \bibinfo
  {author} {\bibfnamefont {M.}~\bibnamefont {de~Simone}}, \bibinfo {author}
  {\bibfnamefont {M.}~\bibnamefont {Coreno}}, \bibinfo {author} {\bibfnamefont
  {A.}~\bibnamefont {Kivim{\"{a}}ki}}, \bibinfo {author} {\bibfnamefont
  {J.}~\bibnamefont {{\AA}hlund}}, \bibinfo {author} {\bibfnamefont
  {L.}~\bibnamefont {Lozzi}}, \bibinfo {author} {\bibfnamefont
  {B.}~\bibnamefont {Brena}}, \ and\ \bibinfo {author} {\bibfnamefont
  {C.}~\bibnamefont {Puglia}},\ }\href {\doibase 10.1002/pssb.201451147}
  {\bibfield  {journal} {\bibinfo  {journal} {Physica Status Solidi (B) Basic
  Research}\ }\textbf {\bibinfo {volume} {252}},\ \bibinfo {pages} {1259}
  (\bibinfo {year} {2015})}\BibitemShut {NoStop}%
\bibitem [{\citenamefont {Kroll}\ \emph {et~al.}(2012)\citenamefont {Kroll},
  \citenamefont {Kraus}, \citenamefont {Sch{\"{o}}nfelder}, \citenamefont
  {Aristov}, \citenamefont {Molodtsova}, \citenamefont {Hoffmann},\ and\
  \citenamefont {Knupfer}}]{45Kroll2012}%
  \BibitemOpen
  \bibfield  {author} {\bibinfo {author} {\bibfnamefont {T.}~\bibnamefont
  {Kroll}}, \bibinfo {author} {\bibfnamefont {R.}~\bibnamefont {Kraus}},
  \bibinfo {author} {\bibfnamefont {R.}~\bibnamefont {Sch{\"{o}}nfelder}},
  \bibinfo {author} {\bibfnamefont {V.~Y.}\ \bibnamefont {Aristov}}, \bibinfo
  {author} {\bibfnamefont {O.~V.}\ \bibnamefont {Molodtsova}}, \bibinfo
  {author} {\bibfnamefont {P.}~\bibnamefont {Hoffmann}}, \ and\ \bibinfo
  {author} {\bibfnamefont {M.}~\bibnamefont {Knupfer}},\ }\href {\doibase
  10.1063/1.4738754} {\bibfield  {journal} {\bibinfo  {journal} {Journal of
  Chemical Physics}\ }\textbf {\bibinfo {volume} {137}} (\bibinfo {year}
  {2012}),\ 10.1063/1.4738754}\BibitemShut {NoStop}%
\bibitem [{\citenamefont {Liao}\ \emph {et~al.}(2005)\citenamefont {Liao},
  \citenamefont {Watts},\ and\ \citenamefont {Huang}}]{46Liao2005}%
  \BibitemOpen
  \bibfield  {author} {\bibinfo {author} {\bibfnamefont {M.~S.}\ \bibnamefont
  {Liao}}, \bibinfo {author} {\bibfnamefont {J.~D.}\ \bibnamefont {Watts}}, \
  and\ \bibinfo {author} {\bibfnamefont {M.~J.}\ \bibnamefont {Huang}},\ }\href
  {\doibase 10.1021/jp0581476} {\bibfield  {journal} {\bibinfo  {journal}
  {Journal of Physical Chemistry A}\ }\textbf {\bibinfo {volume} {109}},\
  \bibinfo {pages} {7988} (\bibinfo {year} {2005})}\BibitemShut {NoStop}%
\bibitem [{\citenamefont {Fern\'andez-Rodr\'{\i}guez}\ \emph
  {et~al.}(2015)\citenamefont {Fern\'andez-Rodr\'{\i}guez}, \citenamefont
  {Toby},\ and\ \citenamefont {van Veenendaal}}]{47Fernandez-Rodriguez2015}%
  \BibitemOpen
  \bibfield  {author} {\bibinfo {author} {\bibfnamefont {J.}~\bibnamefont
  {Fern\'andez-Rodr\'{\i}guez}}, \bibinfo {author} {\bibfnamefont
  {B.}~\bibnamefont {Toby}}, \ and\ \bibinfo {author} {\bibfnamefont
  {M.}~\bibnamefont {van Veenendaal}},\ }\href {\doibase
  10.1103/PhysRevB.91.214427} {\bibfield  {journal} {\bibinfo  {journal} {Phys.
  Rev. B}\ }\textbf {\bibinfo {volume} {91}},\ \bibinfo {pages} {214427}
  (\bibinfo {year} {2015})}\BibitemShut {NoStop}%
\bibitem [{\citenamefont {Sumimoto}\ \emph {et~al.}(2009)\citenamefont
  {Sumimoto}, \citenamefont {Kawashima}, \citenamefont {Hori},\ and\
  \citenamefont {Fujimoto}}]{48Sumimoto2009}%
  \BibitemOpen
  \bibfield  {author} {\bibinfo {author} {\bibfnamefont {M.}~\bibnamefont
  {Sumimoto}}, \bibinfo {author} {\bibfnamefont {Y.}~\bibnamefont {Kawashima}},
  \bibinfo {author} {\bibfnamefont {K.}~\bibnamefont {Hori}}, \ and\ \bibinfo
  {author} {\bibfnamefont {H.}~\bibnamefont {Fujimoto}},\ }\href {\doibase
  10.1039/b823309h} {\bibfield  {journal} {\bibinfo  {journal} {Journal of the
  Chemical Society. Dalton Transactions}\ ,\ \bibinfo {pages} {5737}} (\bibinfo
  {year} {2009})}\BibitemShut {NoStop}%
\bibitem [{\citenamefont {Haldar}\ \emph {et~al.}(2018)\citenamefont {Haldar},
  \citenamefont {Bhandary}, \citenamefont {Vovusha},\ and\ \citenamefont
  {Sanyal}}]{49Haldar2018}%
  \BibitemOpen
  \bibfield  {author} {\bibinfo {author} {\bibfnamefont {S.}~\bibnamefont
  {Haldar}}, \bibinfo {author} {\bibfnamefont {S.}~\bibnamefont {Bhandary}},
  \bibinfo {author} {\bibfnamefont {H.}~\bibnamefont {Vovusha}}, \ and\
  \bibinfo {author} {\bibfnamefont {B.}~\bibnamefont {Sanyal}},\ }\href
  {\doibase 10.1103/PhysRevB.98.085440} {\bibfield  {journal} {\bibinfo
  {journal} {Phys. Rev. B}\ }\textbf {\bibinfo {volume} {98}},\ \bibinfo
  {pages} {085440} (\bibinfo {year} {2018})}\BibitemShut {NoStop}%
\bibitem [{\citenamefont {Wang}\ \emph {et~al.}(2016)\citenamefont {Wang},
  \citenamefont {Hou}, \citenamefont {Zheng},\ and\ \citenamefont
  {Yan}}]{50Wang2016}%
  \BibitemOpen
  \bibfield  {author} {\bibinfo {author} {\bibfnamefont {X.}~\bibnamefont
  {Wang}}, \bibinfo {author} {\bibfnamefont {D.}~\bibnamefont {Hou}}, \bibinfo
  {author} {\bibfnamefont {X.}~\bibnamefont {Zheng}}, \ and\ \bibinfo {author}
  {\bibfnamefont {Y.}~\bibnamefont {Yan}},\ }\href {\doibase 10.1063/1.4939843}
  {\bibfield  {journal} {\bibinfo  {journal} {Journal of Chemical Physics}\
  }\textbf {\bibinfo {volume} {144}} (\bibinfo {year} {2016}),\
  10.1063/1.4939843}\BibitemShut {NoStop}%
\bibitem [{\citenamefont {Hu}\ and\ \citenamefont {Wu}(2013)}]{51Hu2013}%
  \BibitemOpen
  \bibfield  {author} {\bibinfo {author} {\bibfnamefont {J.}~\bibnamefont
  {Hu}}\ and\ \bibinfo {author} {\bibfnamefont {R.}~\bibnamefont {Wu}},\ }\href
  {\doibase 10.1103/PhysRevLett.110.097202} {\bibfield  {journal} {\bibinfo
  {journal} {Physical Review Letters}\ }\textbf {\bibinfo {volume} {110}}
  (\bibinfo {year} {2013}),\ 10.1103/PhysRevLett.110.097202}\BibitemShut
  {NoStop}%
\bibitem [{\citenamefont {Cohen}\ \emph {et~al.}(2012)\citenamefont {Cohen},
  \citenamefont {Mori-S{\'{a}}nchez},\ and\ \citenamefont
  {Yang}}]{52Cohen2012}%
  \BibitemOpen
  \bibfield  {author} {\bibinfo {author} {\bibfnamefont {A.~J.}\ \bibnamefont
  {Cohen}}, \bibinfo {author} {\bibfnamefont {P.}~\bibnamefont
  {Mori-S{\'{a}}nchez}}, \ and\ \bibinfo {author} {\bibfnamefont
  {W.}~\bibnamefont {Yang}},\ }\href {\doibase 10.1021/cr200107z} {\enquote
  {\bibinfo {title} {{Challenges for density functional theory}},}\ } (\bibinfo
  {year} {2012})\BibitemShut {NoStop}%
\bibitem [{\citenamefont {Dale}\ \emph {et~al.}(1968)\citenamefont {Dale},
  \citenamefont {Williams}, \citenamefont {Johnson},\ and\ \citenamefont
  {Thorp}}]{53Dale1968}%
  \BibitemOpen
  \bibfield  {author} {\bibinfo {author} {\bibfnamefont {B.~W.}\ \bibnamefont
  {Dale}}, \bibinfo {author} {\bibfnamefont {R.~J.}\ \bibnamefont {Williams}},
  \bibinfo {author} {\bibfnamefont {C.~E.}\ \bibnamefont {Johnson}}, \ and\
  \bibinfo {author} {\bibfnamefont {T.~L.}\ \bibnamefont {Thorp}},\ }\href
  {\doibase 10.1063/1.1670617} {\bibfield  {journal} {\bibinfo  {journal} {The
  Journal of Chemical Physics}\ }\textbf {\bibinfo {volume} {49}},\ \bibinfo
  {pages} {3441} (\bibinfo {year} {1968})}\BibitemShut {NoStop}%
\bibitem [{\citenamefont {Iv\'ady}\ \emph {et~al.}(2014)\citenamefont
  {Iv\'ady}, \citenamefont {Simon}, \citenamefont {Maze}, \citenamefont
  {Abrikosov},\ and\ \citenamefont {Gali}}]{ZFSVASP}%
  \BibitemOpen
  \bibfield  {author} {\bibinfo {author} {\bibfnamefont {V.}~\bibnamefont
  {Iv\'ady}}, \bibinfo {author} {\bibfnamefont {T.}~\bibnamefont {Simon}},
  \bibinfo {author} {\bibfnamefont {J.~R.}\ \bibnamefont {Maze}}, \bibinfo
  {author} {\bibfnamefont {I.~A.}\ \bibnamefont {Abrikosov}}, \ and\ \bibinfo
  {author} {\bibfnamefont {A.}~\bibnamefont {Gali}},\ }\href {\doibase
  10.1103/PhysRevB.90.235205} {\bibfield  {journal} {\bibinfo  {journal} {Phys.
  Rev. B}\ }\textbf {\bibinfo {volume} {90}},\ \bibinfo {pages} {235205}
  (\bibinfo {year} {2014})}\BibitemShut {NoStop}%
\bibitem [{\citenamefont {Tsukahara}\ \emph {et~al.}(2016)\citenamefont
  {Tsukahara}, \citenamefont {Kawai},\ and\ \citenamefont
  {Takagi}}]{Tsukahara2016}%
  \BibitemOpen
  \bibfield  {author} {\bibinfo {author} {\bibfnamefont {N.}~\bibnamefont
  {Tsukahara}}, \bibinfo {author} {\bibfnamefont {M.}~\bibnamefont {Kawai}}, \
  and\ \bibinfo {author} {\bibfnamefont {N.}~\bibnamefont {Takagi}},\ }\href
  {\doibase 10.1063/1.4940138} {\bibfield  {journal} {\bibinfo  {journal} {The
  Journal of Chemical Physics}\ }\textbf {\bibinfo {volume} {144}},\ \bibinfo
  {pages} {044701} (\bibinfo {year} {2016})}\BibitemShut {NoStop}%
\bibitem [{\citenamefont {Wade}(2000)}]{SOCFe}%
  \BibitemOpen
  \bibfield  {author} {\bibinfo {author} {\bibfnamefont {K.}~\bibnamefont
  {Wade}},\ }\href {\doibase
  10.1002/1099-0739(200008)14:8<449::AID-AOC14>3.0.CO;2-6} {\bibfield
  {journal} {\bibinfo  {journal} {Applied Organometallic Chemistry}\ }\textbf
  {\bibinfo {volume} {14}},\ \bibinfo {pages} {449} (\bibinfo {year}
  {2000})}\BibitemShut {NoStop}%
\bibitem [{\citenamefont {Yu}\ \emph {et~al.}(2016)\citenamefont {Yu},
  \citenamefont {Li},\ and\ \citenamefont {Truhlar}}]{54Yu2016}%
  \BibitemOpen
  \bibfield  {author} {\bibinfo {author} {\bibfnamefont {H.~S.}\ \bibnamefont
  {Yu}}, \bibinfo {author} {\bibfnamefont {S.~L.}\ \bibnamefont {Li}}, \ and\
  \bibinfo {author} {\bibfnamefont {D.~G.}\ \bibnamefont {Truhlar}},\ }\href
  {\doibase 10.1063/1.4963168} {\bibfield  {journal} {\bibinfo  {journal}
  {Journal of Chemical Physics}\ }\textbf {\bibinfo {volume} {145}} (\bibinfo
  {year} {2016}),\ 10.1063/1.4963168}\BibitemShut {NoStop}%
\bibitem [{\citenamefont {Wang}\ \emph
  {et~al.}(2018{\natexlab{a}})\citenamefont {Wang}, \citenamefont {Yang},
  \citenamefont {Ye}, \citenamefont {Zheng},\ and\ \citenamefont
  {Yan}}]{55Wang2018a}%
  \BibitemOpen
  \bibfield  {author} {\bibinfo {author} {\bibfnamefont {X.}~\bibnamefont
  {Wang}}, \bibinfo {author} {\bibfnamefont {L.}~\bibnamefont {Yang}}, \bibinfo
  {author} {\bibfnamefont {L.}~\bibnamefont {Ye}}, \bibinfo {author}
  {\bibfnamefont {X.}~\bibnamefont {Zheng}}, \ and\ \bibinfo {author}
  {\bibfnamefont {Y.}~\bibnamefont {Yan}},\ }\href {\doibase
  10.1021/acs.jpclett.8b00808} {\bibfield  {journal} {\bibinfo  {journal}
  {Journal of Physical Chemistry Letters}\ }\textbf {\bibinfo {volume} {9}},\
  \bibinfo {pages} {2418} (\bibinfo {year} {2018}{\natexlab{a}})}\BibitemShut
  {NoStop}%
\bibitem [{\citenamefont {Wang}\ \emph
  {et~al.}(2018{\natexlab{b}})\citenamefont {Wang}, \citenamefont {Li},
  \citenamefont {Zheng},\ and\ \citenamefont {Yang}}]{56Wang2018}%
  \BibitemOpen
  \bibfield  {author} {\bibinfo {author} {\bibfnamefont {Y.}~\bibnamefont
  {Wang}}, \bibinfo {author} {\bibfnamefont {X.}~\bibnamefont {Li}}, \bibinfo
  {author} {\bibfnamefont {X.}~\bibnamefont {Zheng}}, \ and\ \bibinfo {author}
  {\bibfnamefont {J.}~\bibnamefont {Yang}},\ }\href {\doibase
  10.1039/c8cp05759a} {\bibfield  {journal} {\bibinfo  {journal} {Physical
  Chemistry Chemical Physics}\ }\textbf {\bibinfo {volume} {20}},\ \bibinfo
  {pages} {26396} (\bibinfo {year} {2018}{\natexlab{b}})}\BibitemShut {NoStop}%
\bibitem [{\citenamefont {Momma}\ and\ \citenamefont {Izumi}(2011)}]{60VESTA}%
  \BibitemOpen
  \bibfield  {author} {\bibinfo {author} {\bibfnamefont {K.}~\bibnamefont
  {Momma}}\ and\ \bibinfo {author} {\bibfnamefont {F.}~\bibnamefont {Izumi}},\
  }\href {\doibase 10.1107/S0021889811038970} {\bibfield  {journal} {\bibinfo
  {journal} {Journal of Applied Crystallography}\ }\textbf {\bibinfo {volume}
  {44}},\ \bibinfo {pages} {1272} (\bibinfo {year} {2011})}\BibitemShut
  {NoStop}%
\bibitem [{\citenamefont {Stukowski}(2010)}]{61Ovito}%
  \BibitemOpen
  \bibfield  {author} {\bibinfo {author} {\bibfnamefont {A.}~\bibnamefont
  {Stukowski}},\ }\href {\doibase 10.1088/0965-0393/18/1/015012} {\bibfield
  {journal} {\bibinfo  {journal} {Modelling and Simulation in Materials Science
  and Engineering}\ }\textbf {\bibinfo {volume} {18}} (\bibinfo {year}
  {2010}),\ 10.1088/0965-0393/18/1/015012}\BibitemShut {NoStop}%
\end{thebibliography}%
	
\end{document}